\documentclass[aps,prl,superscriptaddress,onecolumn, showpacs,preprintnumbers,amsmath,amssymb]{revtex4}

\usepackage{longtable}
\usepackage{multirow}
\usepackage{hyperref}
\usepackage{amsmath}
\usepackage{mathtools}
\usepackage{xcolor}
\usepackage{ulem}

\usepackage[pdftex]{graphicx}

\usepackage{color}
\hypersetup{ colorlinks,
linkcolor=blue,
filecolor=black,
urlcolor=black,
citecolor=blue }

\begin{document}

\title{Topological transition in a coupled dynamics in random networks}

\author{P. F. Gomes}
\affiliation{Applied Complex Network Group of Jata\'i, Federal University of Jata\'i, Jata\'i, 75801-615, Brazil}

\author{H. A. Fernandes}
\affiliation{Applied Complex Network Group of Jata\'i, Federal University of Jata\'i, Jata\'i, 75801-615, Brazil}

\author{A. A. Costa}
\affiliation{Applied Complex Network Group of Jata\'i, Federal University of Jata\'i, Jata\'i, 75801-615, Brazil}

\begin{abstract}

In this work, we study the topological transition on the associated networks in a model proposed by Saeedian et al. (Scientific Reports 2019 9:9726), which considers a coupled dynamics of node and link states. Our goal was to better understand the two observed phases, so we use another network structure (the so called random geometric graph - RGG) together with other metrics borrowed from network science. We observed a topological transition on the two associated networks, which are subgraphs of the full network. As the links have two possible states (friendly and non-friendly), we defined each associated network as composed of only one type of link. The (non) friendly associated network has (non) friendly links only. This topological transition was observed from the domain distribution of each associated network between the two phases of the system (absorbing and active). We also showed that another metric from network science called modularity (or assortative coefficient) can also be used as order parameter, giving the same phase diagram as the original order parameter from the seminal work. On the absorbing phase the absolute value of the modularity for each associated network reaches a maximum value, while on the active phase they fall to the minimum value.

\end{abstract}


\maketitle


\section{Introduction}

Social dynamics is the study of the macroscopic properties of human communities determined by the interactions between people within this group. It has been an increasingly active field of research under the area of Statistical Mechanics~\cite{Castellano2009,Galam2012,delVicario2017} and some topics such as opinion dynamics, fake news, rumor spreading, human crowd dynamics, culture evolution, emergence of revolutions and terrorism, social networks and urban agglomerations are becoming even more important nowadays. To study these topics in order to have some clues and estimates of the main properties underlying the interactions, we can make use of models (or propose one) and perform simulations of the dynamics involved on lattices which are used to mimic the society, environment etc. In this field, networks (or graphs) has becoming a paramount asset in the tool box used to study lattice models~\cite{Newman2010,Barabasi2016}. In a network, the nodes represent agents or people and links represent interactions (connections) between pairs of these people. The early models focused on the dynamics of the node states in a fixed network. Since then, many different features have been added to the models based on real-phenomena observations. Properties and dynamics of the network itself have already been vastly explored and can also be coupled with the agent dynamics in a coevolution dynamics~\cite{Blasius2008,Min2017,Reia2020axelrod}. 

Another type of coevolution dynamics is to consider also the state of links (connections) instead of just the state of agents (nodes). So the states of links and agents evolve with time together in an interactive coupled dynamics. In this case the network is static, as there is no links rewiring. Usually, on social dynamics, the interaction between the agents is positive like friendship on a Facebook network. However, in reality this interaction can also be negative: two people in real life may be enemies. So, a given person (or the whole group) can be comfortable or frustrated, happy or unhappy, balanced or unbalanced, depending on his/her link and node states. This is the idea behind the Heider's concept of balanced and unbalanced triadic groups (triangles) of individuals~\cite{Heider1946,Heider1964}, which is the base of the structural balance theory~\cite{Marvel2009,Traag2013}. Outside this triadic structure, the link states have also been studied in a regular network where the iteration with the agent states are responsible to determine the time evolution of the system~\cite{Gracia2012,Carro2014,Saeedian2019,Saeedian2020,Pham2020}.

Many of the models studied on social dynamics are classified as non-equilibrium models in statistical mechanics and one hallmark of such systems is the presence of, at least, one absorbing phase~\cite{Hinrichsen2000} which is a phase where the system gets trapped and cannot escape from it. The other phase is usually called ``active'' phase which is a phase where the kinetics of the model occurs. There is no general theory of non-equilibrium phase transitions like in equilibrium ones, so absorbing phase transitions have been studied in different models with different approaches. In this sense, network science has been increasingly more important as a tool to fully describe the different phenomena obtained from real-world observation involving non-equilibrium models~\cite{Costa2011,Arruda2017a,ACosta2013,Reia2017d,Pazzini2021,Fernandes2016,Brito2020}. 

One important network feature is its component distribution. Component is a connected subgraph of the network. On the other hand a cultural domain is more specific, defined as a connected subgraph with the same attribute. This concept has its motivations from studies of cultural dissemination, using for example the Axelrod's model~\cite{Axelrod1997,Hernandez2018c}. The cultural domains have been used as the order parameter in the study of cultural dissemination ~\cite{Klemm2003,Klemma2003b,Reia2016,Reia2020axelrod}. 

Other more elaborated network feature of a network is the mixing pattern, which is defined by how often one given node is connected to a node of the same attribute~\cite{Newman2002}, which again can be the cultural state of the node. The idea is to quantify the feature known as homophily or assortative mixing~\cite{Avin2020}, which is the tendency one individual has to associate to a similar one rather than a different one. For an example, homophily is one of the features in the Axelrod's model~\cite{Axelrod1997,Reia2020c}, where the probability of interaction between two individuals is proportional to their cultural similarity. The opposite is called disassortativity mixing, and the most familiar example of this is mixing by gender in sexual contact networks: the majority of sexual partnerships are between individuals of opposite sex~\cite{Newman2010}. Modularity or assortativity coefficient has been successfully used as a measure of the mixing patterns on a given network~\cite{Newman2003} and in the study of explosive synchronization in a complex neuronal network~\cite{Roy2021}. 

In this work, we build up on the model developed by Saeedian \textit{et. al}~\cite{Saeedian2019} in order study the topological transition of the associated networks between the absorbing and active phases of the system. By associated network we mean the sub-graph of the full network containing only one type of links, as it will be properly defined together with components and domains. The full network is static and only the associated networks evolve with time as the links changes its states. We use different measures borrowed from network science, such as components and cultural domains, to evaluate an intuitive picture of this topological transition between the two phases when the model is implemented in a random geometric graph (RGG). We also employ the modularity for each link state as an order parameter of the system and obtained the phase diagram. 

\section{Methods}

\subsection{The model}
\begin{figure} [ht!]
\centering
\includegraphics[width=4.4 in]{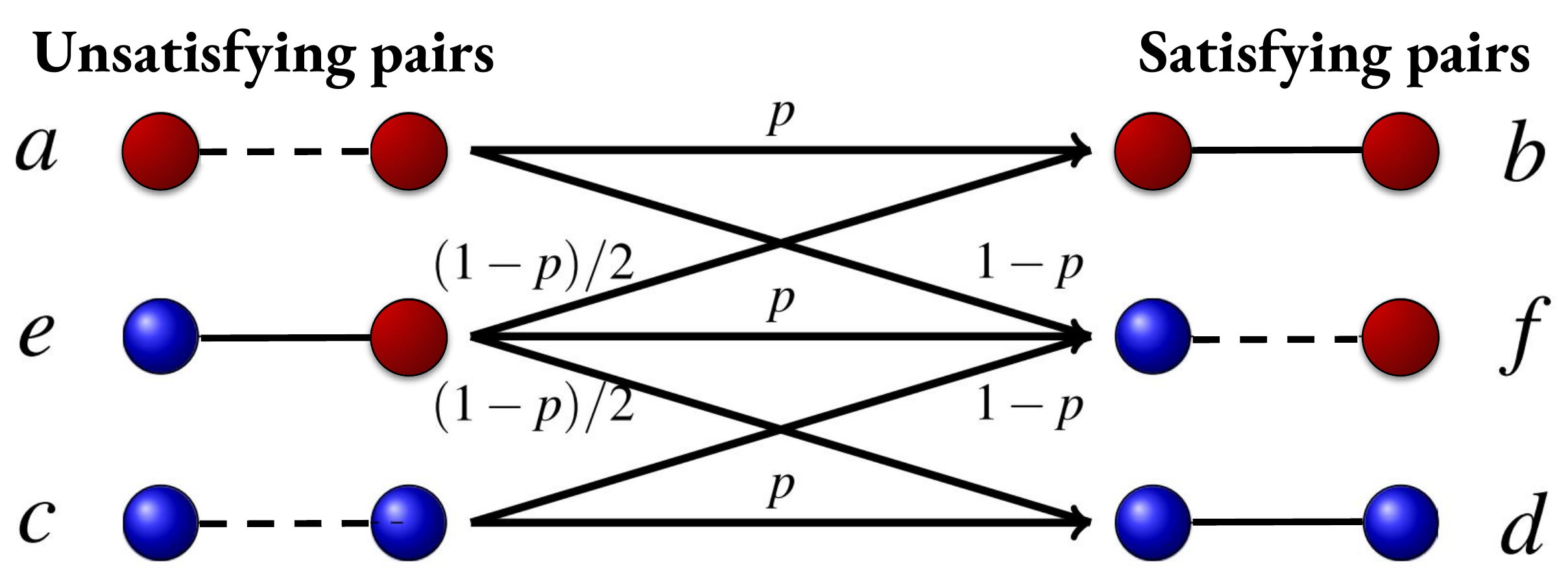}
\caption{Six possible types of pairs based on two types of link states, friendly (solid line) and non-friendly (dashed line), and two types of node states opinion, blue and red. The pairs $a, c$ and $e$ are unsatisfying while pairs $b, d$ and $f$ are satisfying. $p$ is the probability of link update and $1 - p$ is the probability of node update. Adapted from Ref.~\cite{Saeedian2019}.} \label{Dinamica}
\end{figure}

The model is composed of $M$ nodes (or agents) and $L$ links. Each node can have two different cultural (or opinion) states: red and blue. In the same way, each link (or connection) between two nodes also has two possible states: friendly or non-friendly. Then there are six different pair combinations of agents and links, categorized as satisfying and unsatisfying combinations, as depicted in Fig. \ref{Dinamica}. One pair is said satisfying if two nodes of the same type are connected by a friendly link (solid line in the figure), or if a pair of nodes with different states is connected by an unfriendly link (dashed line). Otherwise the pair is unsatisfying. The idea, as in Axelrod's model, is that one node likes to be friend with another node of the same state and do not want to be friend with a node of different state.

Both nodes and links have a specific dynamics for evolving their states with time, and the control parameter of the model is the probability $p$. It dictates the probabilities of how unsatisfying pairs become satisfying as indicated in Fig. \ref{Dinamica}. During the evolution of the system, the time is measured in Monte Carlo steps (MCS). In each MCS, $L$ pairs are randomly selected in sequence and analyzed. If the selected pair is satisfying ($b$, $f$ or $d$ in Fig. \ref{Dinamica}) nothing happens. Otherwise if it is unsatisfying ($a$, $e$ or $c$), it becomes satisfying by changing the link state or the state of one of the nodes. For the three unsatisfying cases, $p$ is the chance for changing the link states. The types $a$ and $c$ can change one-node state with chance $1-p$ and the type $e$, on its turn, can also change the blue node to red or the red one to blue, both events with chance $(1-p)/2$. It is worth mentioning that, even though each unsatisfying selected pair always become satisfying, its nodes are possibly paired with other nodes on the network, and these other pairs could be satisfying and eventually become unsatisfying after the change.

The order parameter of the model is the density of the $e$-type pair ($\rho_e$), as the other two unsatisfying densities (types $a$ and $c$) are proportional to $\rho_e$ in the stationary state $t \rightarrow \infty$, from mean-field approximation results~\cite{Saeedian2019}. So if $\rho_e=0$ the total unsatisfying density is zero and the dynamics comes to a halt, which means that the system is in an absorbing state and all pairs are satisfying.

On the other hand, in the active state there is a constant change of pairs from unsatisfying to satisfying and vice-versa. The unsatisfying pair becomes satisfying during the dynamics when another satisfying pair containing one of the nodes of the first pair can become unsatisfying preventing the system to reach the absorbing state.

\subsection{Random Geometric Graph: RGG}

We use a random geometric graph (RGG) as the connectivity pattern between the nodes~\cite{Dall2002,penrose2003}. The RGG is constructed by placing $M$ nodes at random in a square box with size $z$. The criterion to define the connectivity is the following: two sites are connected, i.e., they are neighbors, if and only if the euclidean distance between them is less than $r$. We maintain the superficial density of sites $w=M/z^2$ constant in order to discard its effect when the size $M$ of the network is adjusted. So, we choose the linear size of the square box as $\sqrt{M/w}$. Since the effective area of one site is $1/w$, we adopt the linear size $1/\sqrt{w}$ as the spatial scale to measure all distances in our study~\cite{Gomes2019}. In this way, the network properties do not depend on the linear size of the square box and are uniquely defined by the radius $r$ measured in units of $1/\sqrt{w}$. As any constant value for $w$ works, we set $w=1.0$ so the linear size of the square box becomes $\sqrt{M}$, and the radius is a dimensionless number. In this way, the average degree $K$ of the network varies from $K=0$, for $r=0$, to $K=M-1$, when $r \geq z \sqrt{2} /2$ (half the diagonal of the square). This random geometric graph is the originally random graph introduced to model wireless communication networks~\cite{Gilbert1961} and it has been recently used with different applications~\cite{Estrada2016a,Reia2019welfarism,Reia2020axelrod,Vilela2020}.

\subsection{Computational details}

The model is simulated on surfaces with $M=16^2$, $32^2$, $64^2$ and $128^2$ nodes on a square box with linear size $\sqrt{M}$ with periodic boundary conditions. See Appendix~\ref{appendixA} for a list and meanings of the used symbols. The steps of the algorithm are as follows:
\begin{enumerate}
\item Definition of the input parameters: $M$, $p$, $B$, $Q$, $r$, $s$ $\phi_0$ and $\ell_0$.
\item The network is built and the number $L$ of links (pairs of connected nodes) are calculated.
\item The population at $t=0$ is created where $\phi_0$ is the chance of a given node to be red and $\ell_0$ is the chance of a given link to be friendly.
\item In the simulation, each random selected link has its states evaluated according to the described rules considering the probability $p$. One Monte Carlo step (MCS) means that $L$ random links were evaluated. 
\item The simulation is performed through $Q$ Monte Carlo steps (MCS) in order to let the densities stabilize on their steady state values. Even though the system may also end up in an absorbing state on the active regime due to fluctuations. However we did not observed this feature in the time scales analyzed here.
\item After this $Q$ MCS, the simulation continues for $B$ more MCS, so that the interesting quantities (density, modularities, domains and components) are averaged over these $B$ MCS.
\item For the error calculations, all the previous steps is repeated $D$ times, each one with a different random sample of the RGG network.
\end{enumerate} 
All results shown in this work were obtained after $Q=2000$ MCS. The data was generated using Fortran and the analysis and graphics were developed using Python (packages: Numpy~\cite{Harris2020}, Matplotlib~\cite{Hunter2007}, NetworkX~\cite{Hagberg2008} and Pandas~\cite{McKinney2010}). All codes and data are available upon request. 

\subsection{Analysis}

\subsubsection{Phase diagrams}
All features of the RGG, as the average degree $K$, are defined by the radius $r$ and the network size $M$. In this way, the two control parameters of the model are the probability $p$ and the radius $r$, so that the phase diagram of the system is on the space $r,p$ with the limitation $r>0$ and $0<p<1.0$.

One result from the mean-field approximation is that the critical value $p_c$ is a function of the average degree $K$~\cite{Saeedian2019}:
\begin{equation}
    p_c (K) = 1 - \dfrac{3}{2K-2}, \label{criticallinered}
\end{equation}
where $K=K(r)$ in our case. So, in the phase diagram, each point $K(r)$, $p$ defines if the system is in the active state $p<p_c(K)$ or in the absorbing state otherwise. Even though Eq. (\ref{criticallinered}) is only valid in the thermodynamic limit $M\rightarrow \infty$ for a fully connected network (different then the case studied here), it will be a guide in the analysis of our results.

\subsubsection{Components and domains}

\begin{figure} [ht!] 
\centering
\includegraphics[width=2.1 in]{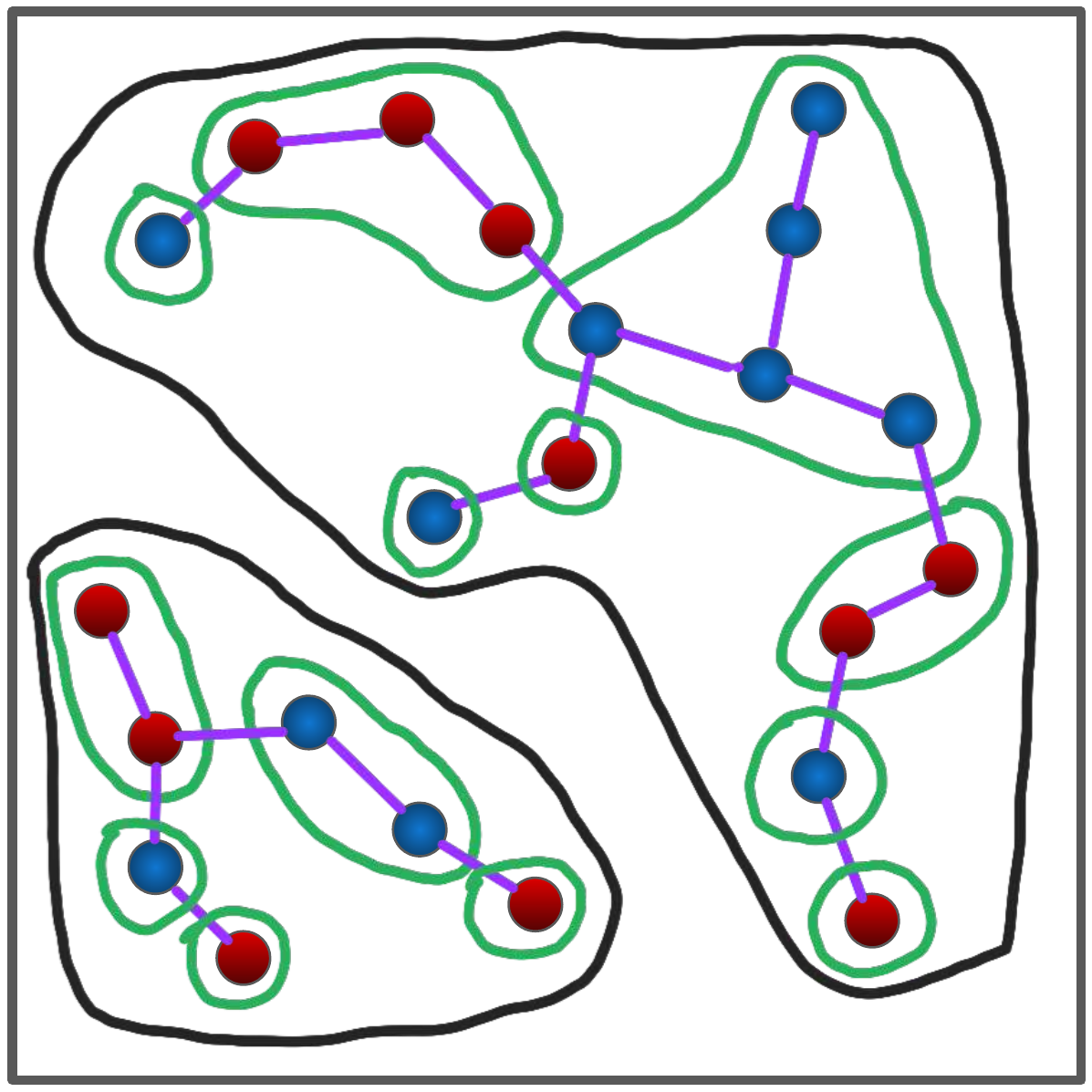}
\caption{Illustration of components and cultural domains considering only one type of link: friendly. The color of the nodes indicates their states and the solid gray lines indicate the links. There are two components $\mathcal{N}_c=2$ indicated by the thick black line. The largest component has $\mathcal{S}_c=15$ nodes. There are $\mathcal{N}_d=13$ cultural domains (indicated by the green lines): five of them with two or more nodes. The other eight domains have only one node. The largest domain has $\mathcal{S}_d =5$ nodes. The total number of nodes is $M_a=22$.} 
\label{comp_dominios4}
\end{figure}

We use the components and cultural domains of each associated network as a way to further characterize the absorbing and active phases. A component to a given network is a group of connected nodes so that it is possible to go from any node to any other node through the connections~\cite{Newman2010}, i.e., a connected subgraph. The size of a component is the number of nodes on it (see Fig.~\ref{comp_dominios4}) and, of course, one isolated node is a component of size 1.

A group of connected nodes with the same attribute~\cite{Han2020a}, as the cultural state (red or blue in this work), is defined as a domain. Therefore one domain is a group of same state nodes inside a component, and there can be more than one domain belonging to the same component. Then the number of domains is always larger or equal than the number of components, as shown in Fig.~\ref{comp_dominios4}. In other words, the size of a domain is always smaller or equal do the size of the component that contains it.

The associated networks are the ones containing only one type of link: friendly network with only friendly links, and non-friendly network containing only non-friendly links. The related quantities are: number of components $\mathcal{N}_c$, number of domains $\mathcal{N}_d$, size of the largest component $\mathcal{S}_c$ and size of the largest domain $\mathcal{S}_d$. In order to compare the results for different network sizes $M$, we use the normalized version of these quantities for each associated network:
\begin{eqnarray}
    n_c &=& \dfrac{\mathcal{N}_c}{M_a}, \qquad \qquad n_d = \dfrac{\mathcal{N}_d}{M_a}, \nonumber \\
        s_c &=& \dfrac{\mathcal{S}_c}{M_a}, \qquad \qquad s_d = \dfrac{\mathcal{S}_d}{M_a}, \nonumber 
\end{eqnarray}
where $M_a$ is the number of nodes of the associated network (friendly or non-friendly). From now on, every time we mention these quantities we refer to the normalized ones.

\subsubsection{Modularity or assortativity coefficient}

We use the assortativity coefficient of each associate network as an alternative order parameter of the system. It is defined as (Eq. (7.69) on Ref.~\cite{Newman2010}):
\begin{equation}
\Lambda = \dfrac{1}{2L} \sum_{ij} \left( A_{ij} - \dfrac{\kappa_i \kappa_j}{2L} \right) \delta (c_i,c_j), \label{lkjweriuowieur}
\end{equation}
where $\delta (a,b)$ is the Kronecker delta, $\kappa_i$ is the degree of node $i$ and $c_i$ is the state of the node $i$ on the associated network (friendly or non-friendly). Here we have: $c_i=1$ (red) and $c_i=2$ (blue). $A_{ij}$ is the adjacency matrix for the associated network, it gives the connection distribution of the network. As we are dealing with simple unweighted and undirected networks, the definition is: $A_{ij} = 1$ if nodes i and j are connected, and zero if they are not connected. The modularity is positive if there are more edges between nodes of the same type than expected by chance, and negative otherwise. In the limit cases, $\Lambda$ is maximum if the network only contains pairs of nodes with the same state and minimum (negative) if it only contains pairs with different node states.

The (full) network is composed of friendly and non-friendly links. We implemented the separation of the friendly and the non-friendly associated networks in order to calculate the modularities for each link type and we call $\Lambda_f$ and $\Lambda_n$ the modularities of the friendly and non-friendly associate networks, respectively. On each case (friendly and non-friendly), the parameters of Eq. (2) refer to the associated network only. For example, for the friendly modularity, $A_{ij}$ is 1 if nodes i and j are connected by a friendly link and zero otherwise, $L$ is the number of friendly links, $\kappa_i$ is the number of friendly links of node i. Any isolated nodes did not enter in the associated networks.

\section{Results}
\subsection{RGG properties}

\begin{figure} [ht!] 
\centering
\includegraphics[width=3.1 in]{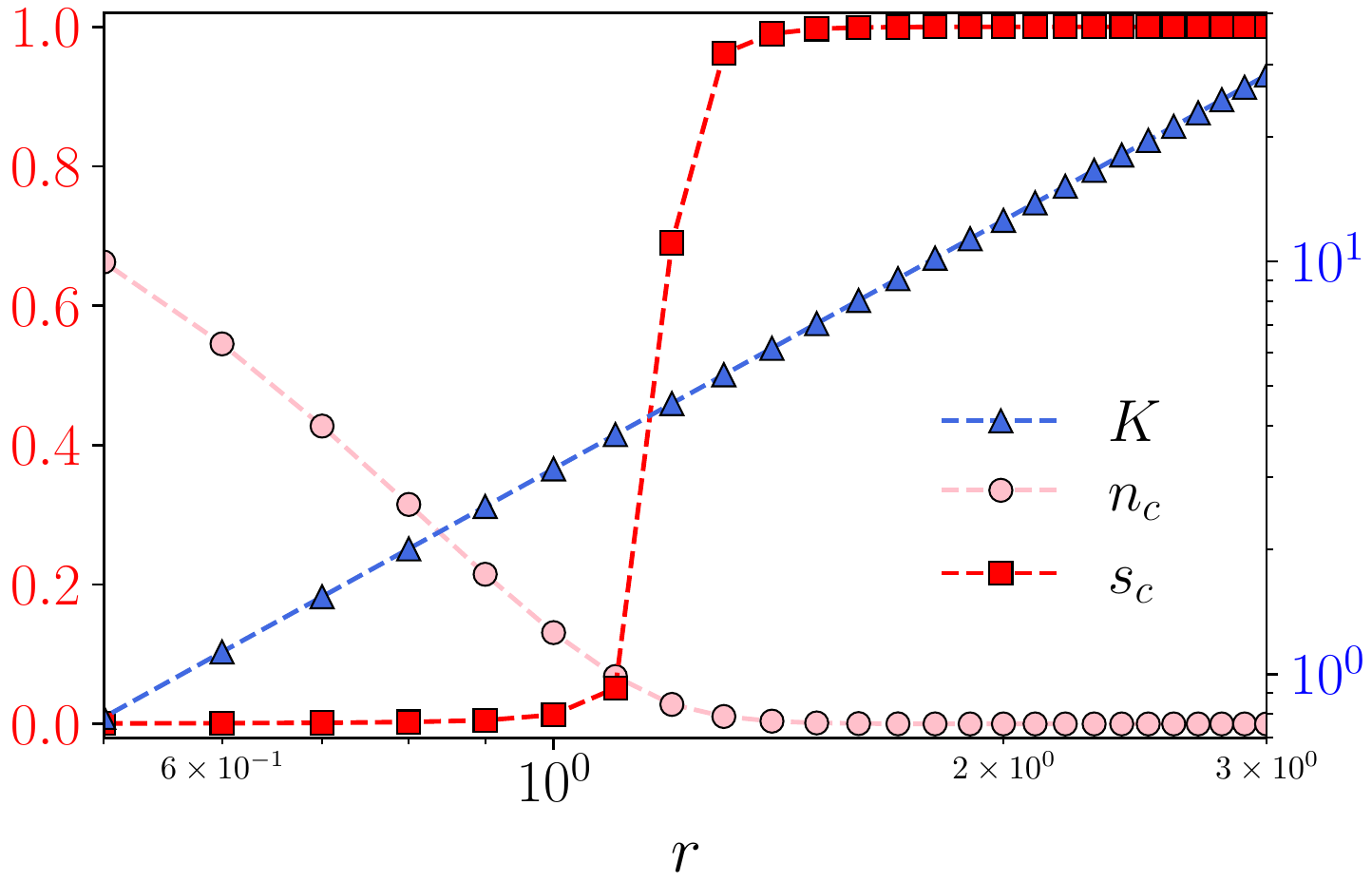}
\caption{Topological properties of the RGG network in the range $0.5<r<3.0$ for $M=128^2$. The network average degree $K$ is the blue triangles (right axes), the number of components $n_c$ is represented by pink circles and the largest component $s_c$ is the red squares (both on the left axes). The dashed lines are just a guide to the eye. As long as $M>K$, the average degree $K$ does not depend on $M$ because we fix $\sqrt{M}$ as the linear size of the square box in the RGG implementation. The data is the average for 15,000 instances of RGGs, and the error bars are negligible.} 
\label{topological_RGG}
\end{figure}

As presented above, the connectivity in the RGG network is controlled by its radius $r$~\cite{Vilela2020}. In this work, in order to look into the phase transition of the model, we set the region of interest as the range $0.5<r<3.0$.
As shown in Fig. \ref{topological_RGG}, the network average degree $K$ goes linear in a log-log scale with $r$ behavior. This figure also shows that the network percolates for $r>1.0$, i.e., there is a single network component~\cite{Reia2020axelrod}, which means $s_c \rightarrow 1$ and $n_c \rightarrow 1/M$.

\subsection{Initial condition}

\begin{figure}[h!]
\centering
\includegraphics[width=3.0 in]{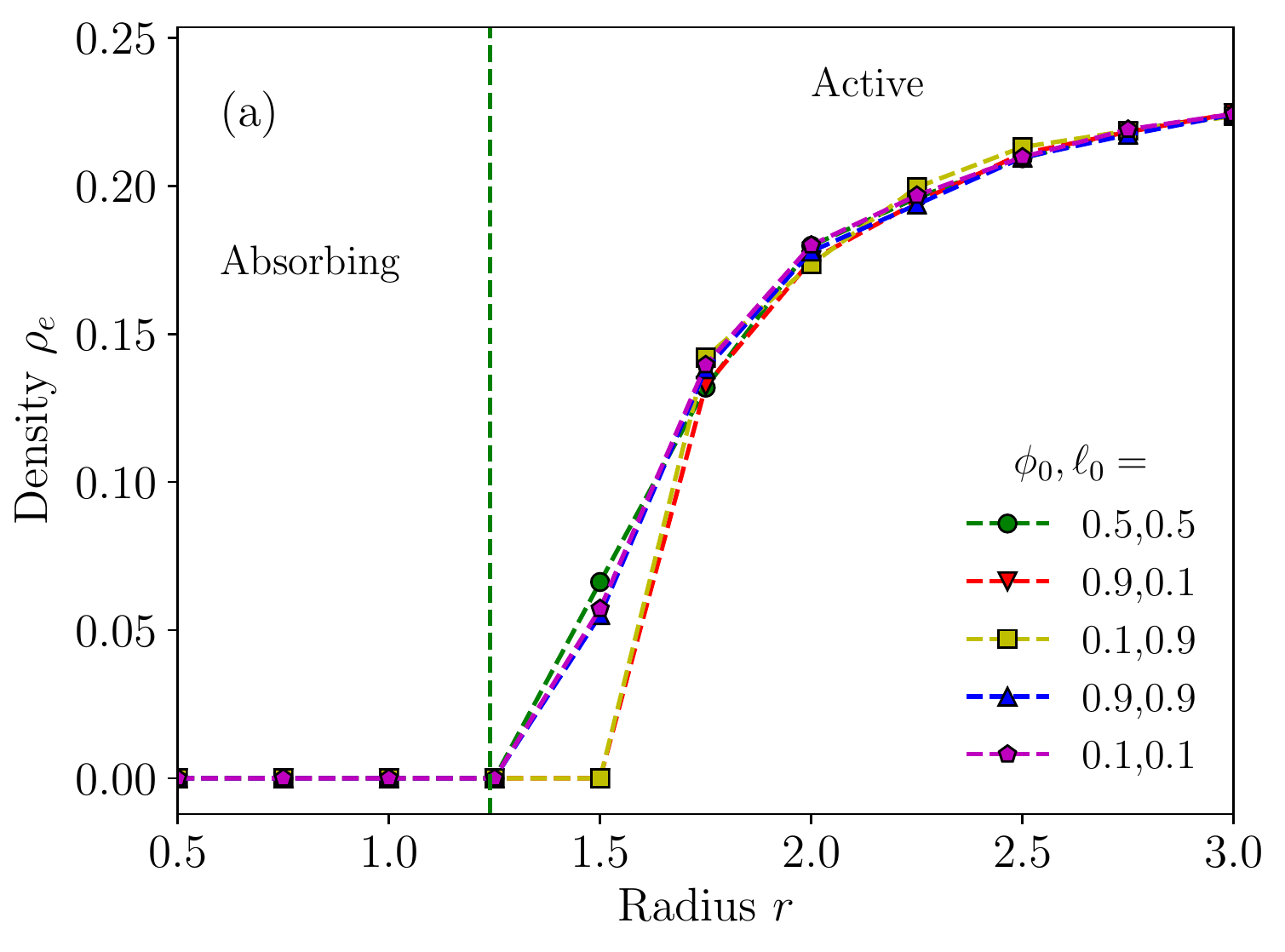}
\includegraphics[width=3.0 in]{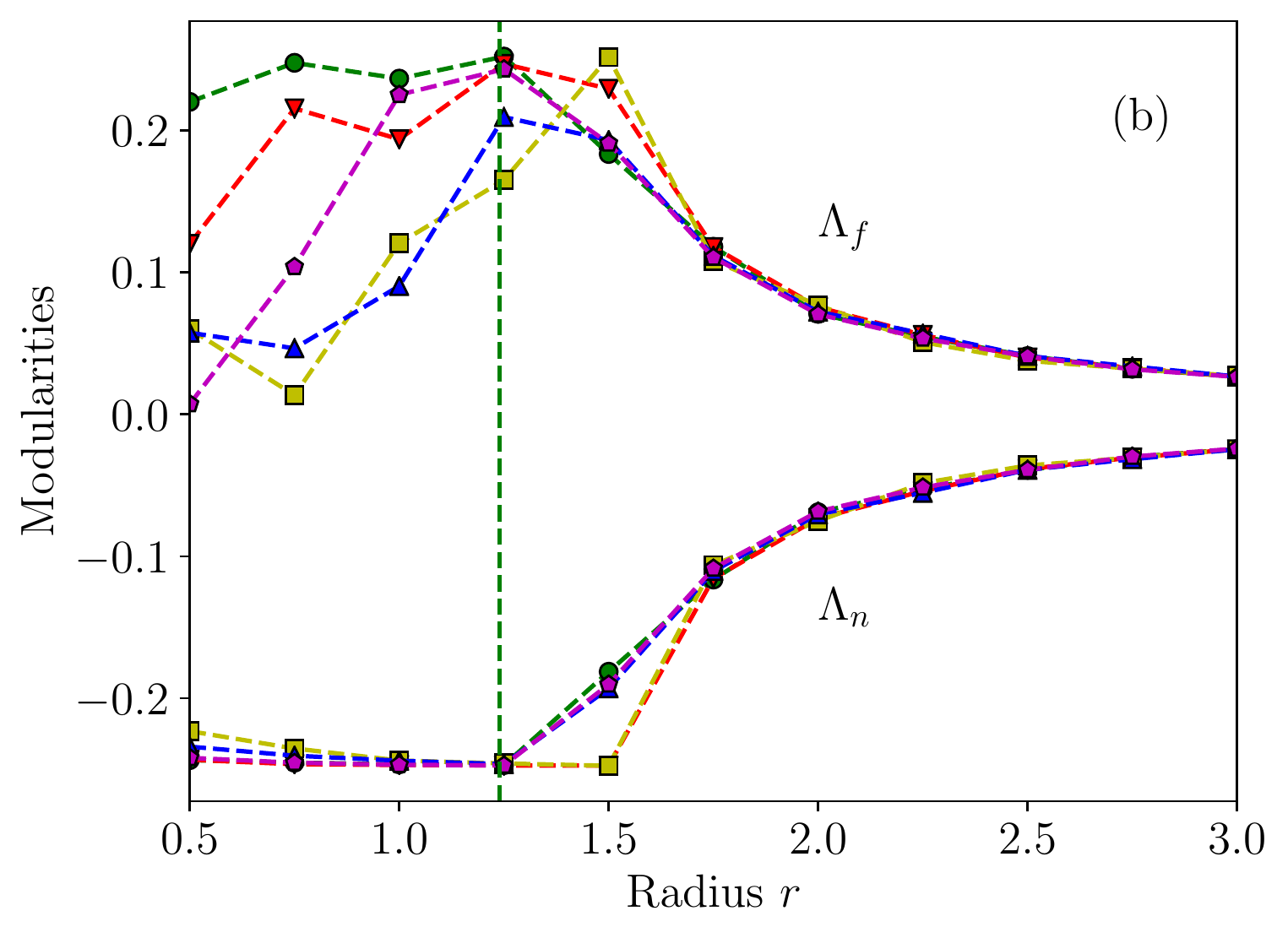}
\caption{Influence of the initial conditions on the order parameters at the stationary regime. The considered parameters are the following: $M=16^2$, $p=0.6$ and each point is an average over $B=2000$ MCS on the stationary regime for a unique instance of the RGG network. Vertical dashed green line indicates the critical value $r_c$ such that $K_c = K(r_c)$ and $p_c=0.6$ satisfies Eq. (\ref{criticallinered}). On this sparse grid the absorbing phase occurs for $r \leq r_c$ and the active one for $r>r_c$ (as indicated). (a) Density $\rho_e$. (b) Modularities $\Lambda_f$ (upper curves) and $\Lambda_n$ (lower curves).} 
\label{fig_grafico_seq64_rls_a1}
\end{figure}

The input parameters $\phi_0$ and $\ell_0$ define the chances of having a red node and a friendly link in the population at initial condition $t=0$. Fig. \ref{fig_grafico_seq64_rls_a1} shows the order parameters in the stationary regime for different combinations of $\phi_0$ and $\ell_0$. The absorbing state is reached regardless of the values of the initial parameters as shown by the behavior of the density $\rho_e$. The non-friendly modularity only shows a small variation at low radius.On the other hand the friendly modularity $\Lambda_f$ shows large oscillations in the absorbing phase. Even though our simulations do not follow the conditions for Eq. (\ref{criticallinered}) to valid, the mean-field result serves well as a indication of the phase transition in these data. From now on, all following results use the initial parameters $\phi_0=\ell_0=0.5$.

\subsection{Phase transition}

\begin{figure}
\centering
\includegraphics[width=3.0 in]{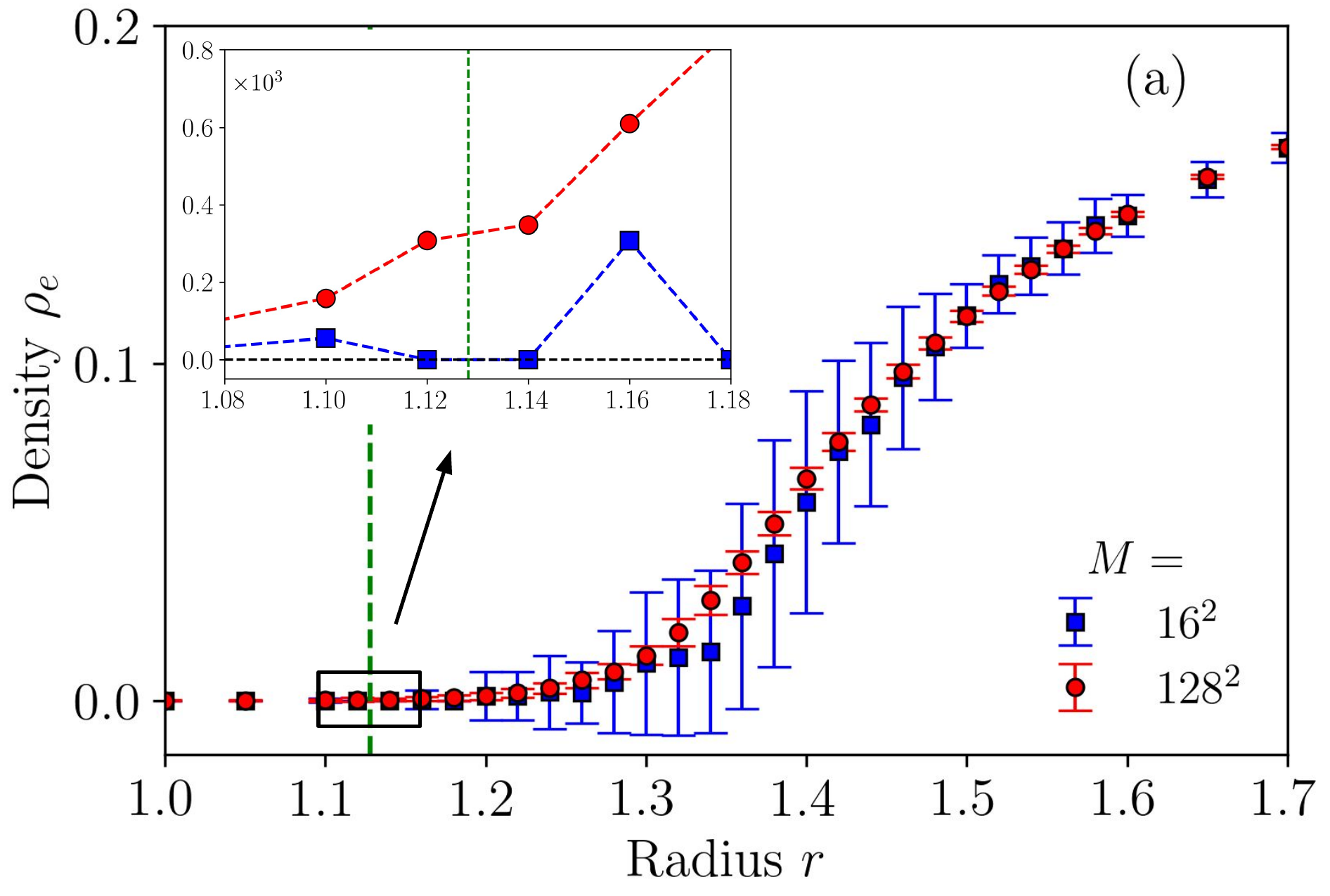}
\includegraphics[width=3.2 in]{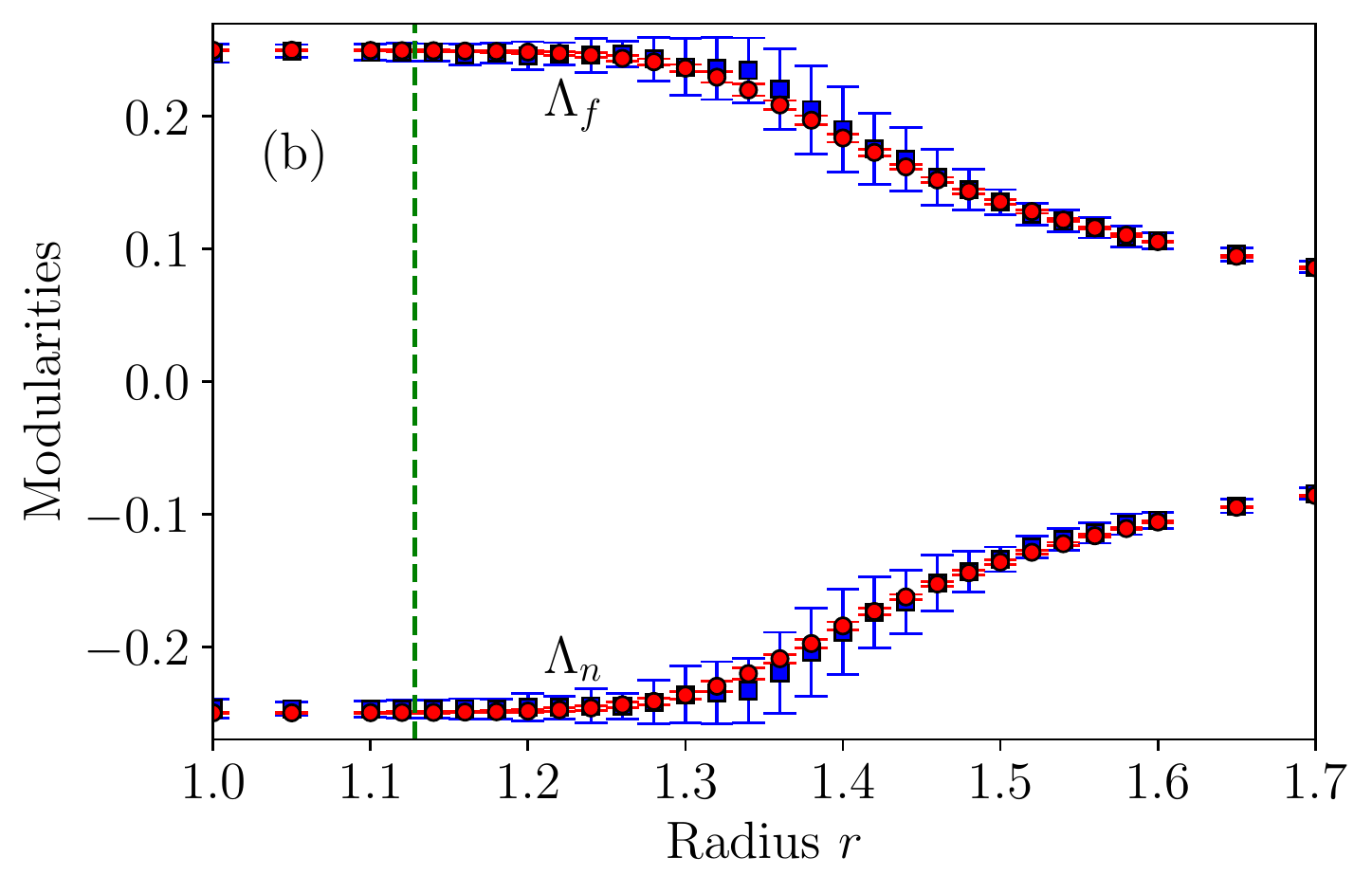}
\includegraphics[width=3.1 in]{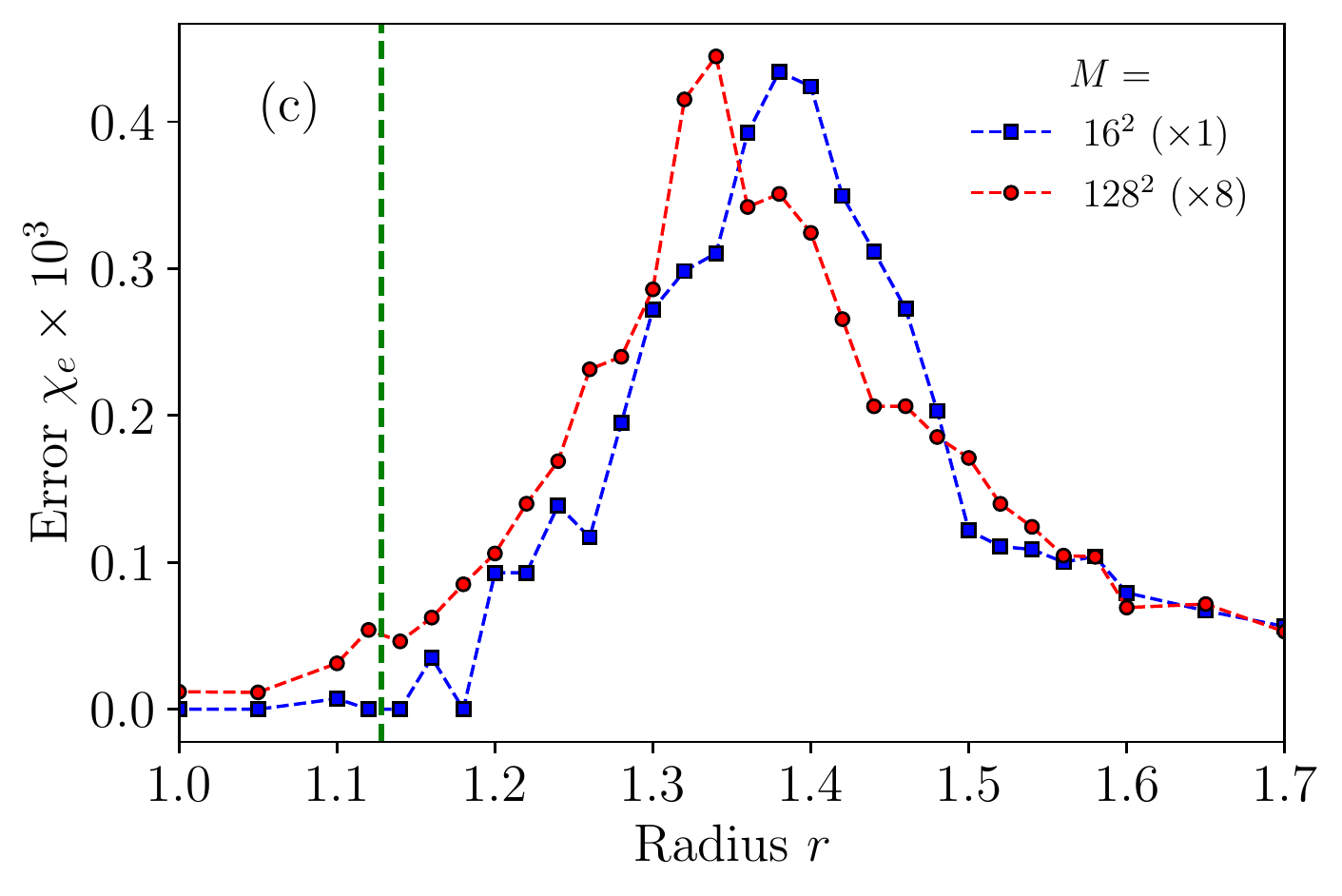}
\includegraphics[width=3.1 in]{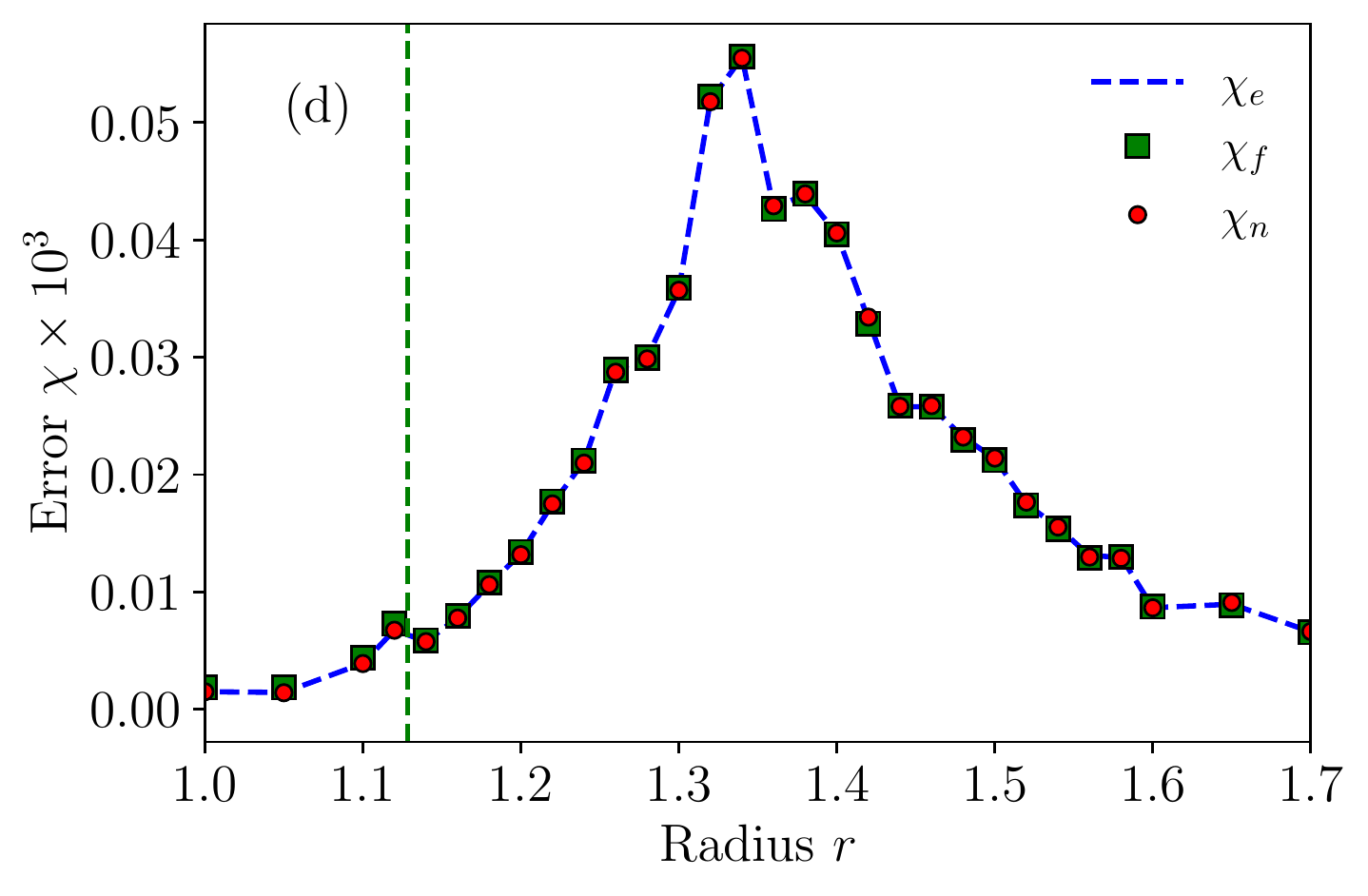}
\caption{Behavior of the order parameters as function of $r$ for the parameters: $p=0.5$, $\phi_0=\ell_0=0.5$ and $B=1000$. Blue square and red circle indicate networks with $M=16^2$ and $M=128^2$ nodes, respectively. The error bars indicate the standard deviation instead of the standard error for better visualization. It was calculated using $D=10^2$ different samples of the RGG network. The vertical dashed green line indicates the critical value of $r$, $r_c \approx 1.128$, such that $K_c=K(r_c)$ and $p=p_c=0.5$ satisfy Eq. (\ref{criticallinered}). (a) Density $\rho_e$. Inset: zoom on the indicated rectangular area: the vertical axis is multiplied by $10^3$ for better visualization and the horizontal dashed line means $\rho_e=0.0$. (b) Modularities $\Lambda_f$ and $\Lambda_n$. (c) Error $\chi_e$ on the density $\rho_e$. Data for the largest size are multiplied by eight. (d) Errors for all 3 order parameters for the largest network size.}
\label{phase_transition_parameters}
\end{figure}

Figs.~\ref{phase_transition_parameters}(a) and (b) shows the behavior of the order parameters $\rho_e$, $\Lambda_f$ and $\Lambda_n$ surrounding the phase transition for different number $M$ of nodes. In the absorbing phase (at low $r$) we have $\rho_e \approx 0.0$ indicating a presence of only satisfying pairs. The modularity $\Lambda_f$ reaches its maximum value in this absorbing phase indicating that all links connect nodes with the same state. In the same way $\Lambda_n$ reaches its minimum value indicating that all the non-friendly links connect nodes with different states. On the other hand the active phase occurs for higher $r$ values, for which unsatisfying pairs start to appear; then $\rho_e>0.0$, $\vert \Lambda_f \vert$ and $\vert \Lambda_n \vert$ decrease to its minimum value.  Moreover, the behavior of the order parameter $\rho_e$ in Fig. \ref{phase_transition_parameters}(a) gives us a clue that the transition between the absorbing and active phase is continuous.

For low values of $r$, and consequently a low average degree $K$ (see Fig. \ref{topological_RGG}), if one node changes its state there are few other pairs of this node that can go back from satisfying to unsatisfying. Thus the absorbing phase can be reached for lower values of $r$ more easily than for large $r$. We can see this effect directly on Fig. \ref{phase_transition_parameters}(a).

The behaviour of $\rho_e$ shown in the inset on Fig.~\ref{phase_transition_parameters}(a) is not in agreement with the mean-field calculation (which is just an approximation here), since the order parameter $\rho_e$ starts to increase in our simulations even in the mean-field absorbing-phase region. Moreover, the fluctuation $\chi_e$ of the order parameter $\rho_e$ present a sharp maximum for $r>r_c$, as shown on Fig. \ref{phase_transition_parameters}(c). Additionally, this maximum shifts to a lower $r$ when the size $M$ is increased: $\approx 1.382$ and $\approx 1.34$ for the sizes $16^2$ and $128^2$. This feature is in agreement with the fact that $r_c$ is valid in the thermodynamic limit. For the largest network size, the modularities $\Lambda_f$ and $\Lambda_n$ give almost exactly the same errors, as shown on Fig. \ref{phase_transition_parameters}(d).

Thus, a qualitative definition of the phase transition is when the first unsatisfying pair appears, which can be difficult to identify due to the numerical fluctuation of the order parameters. Anyway this criterion would lead to a radius around the mean-field approximation result $r_c$. However, a quantitative criterion is the maximum of the standard error (fluctuation) of the order parameters.

\subsection{Phase diagram}

\begin{figure} 
\centering
\includegraphics[width=2.2 in]{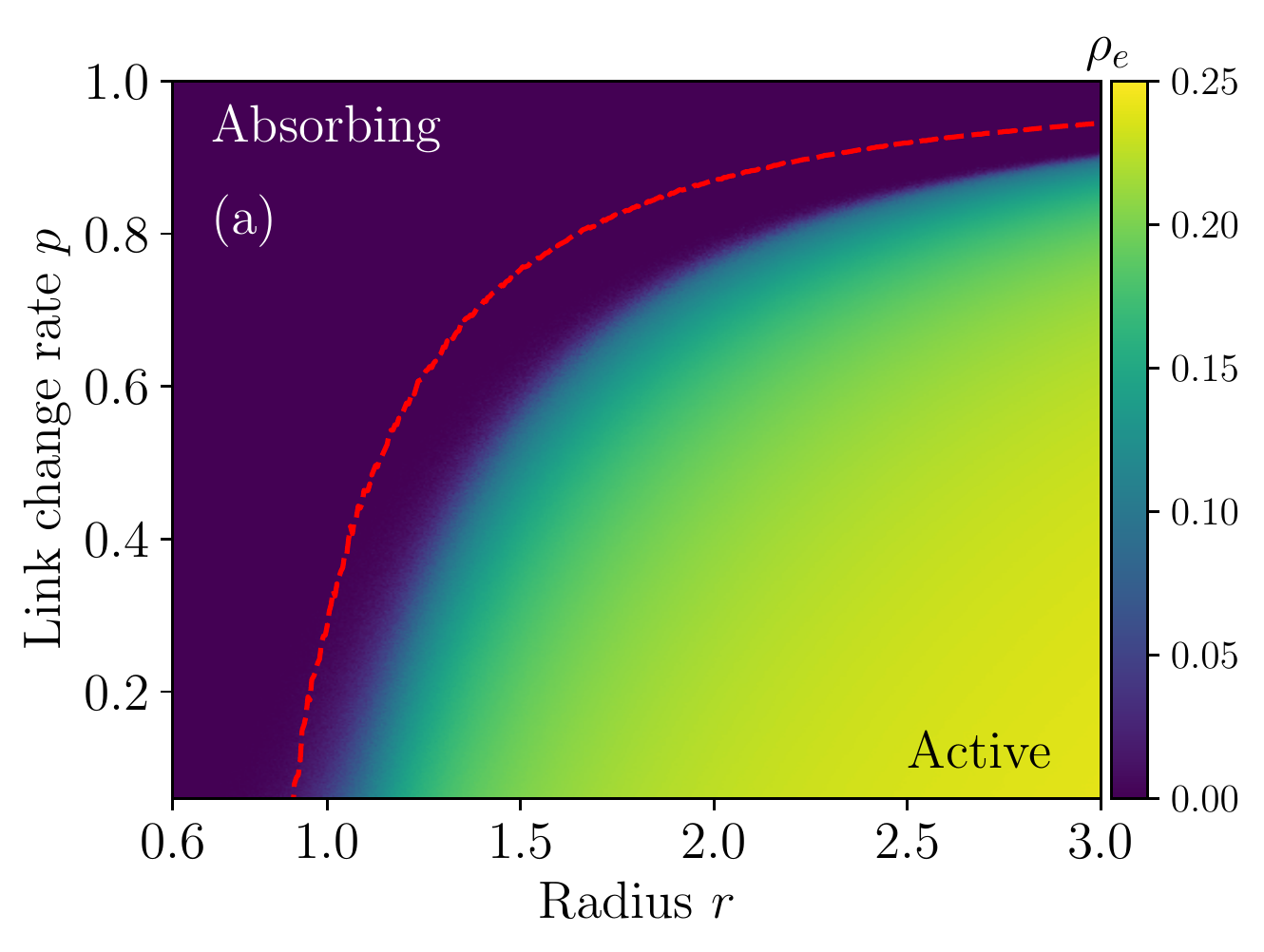}
\includegraphics[width=2.2 in]{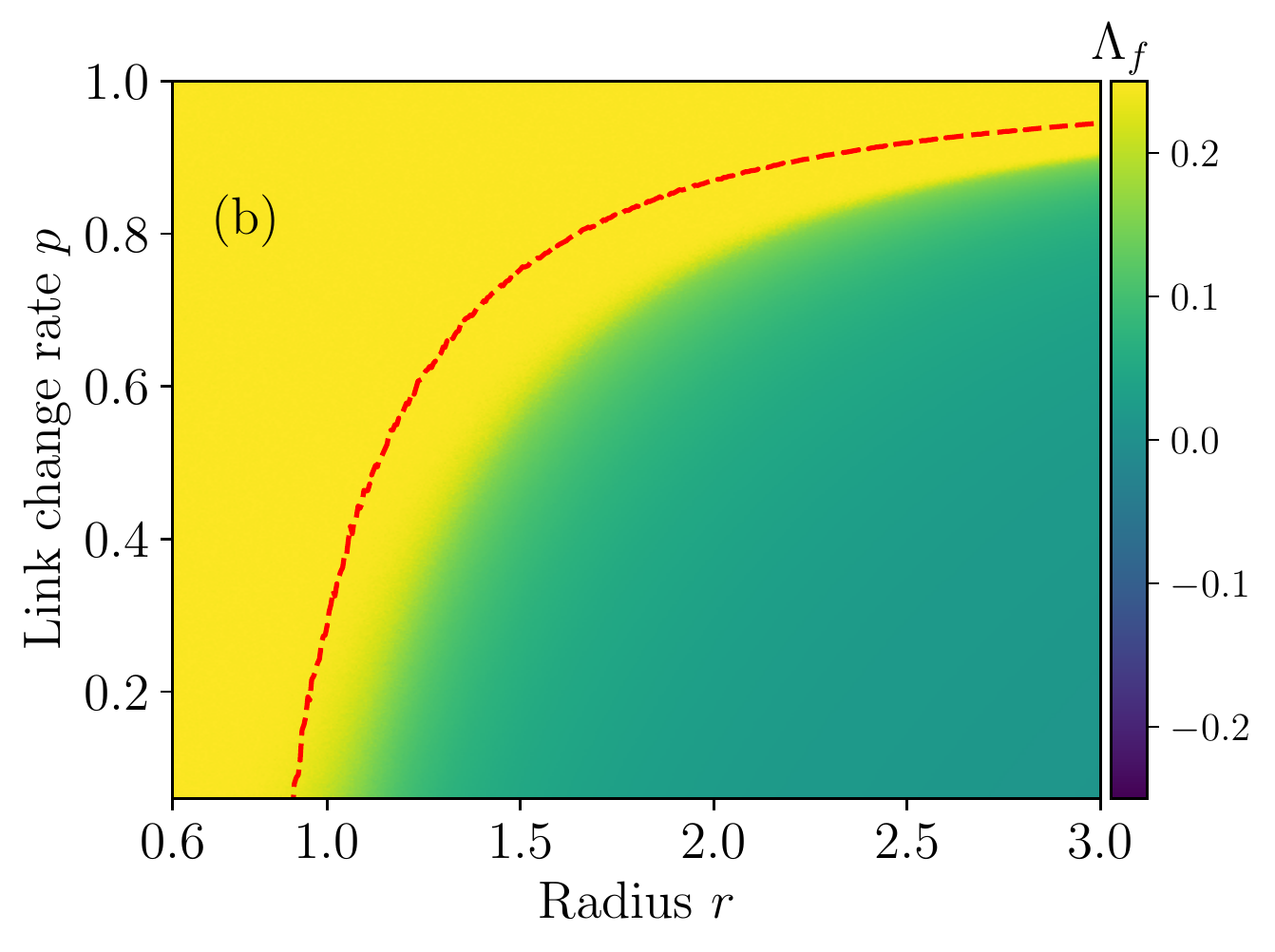}
\includegraphics[width=2.2 in]{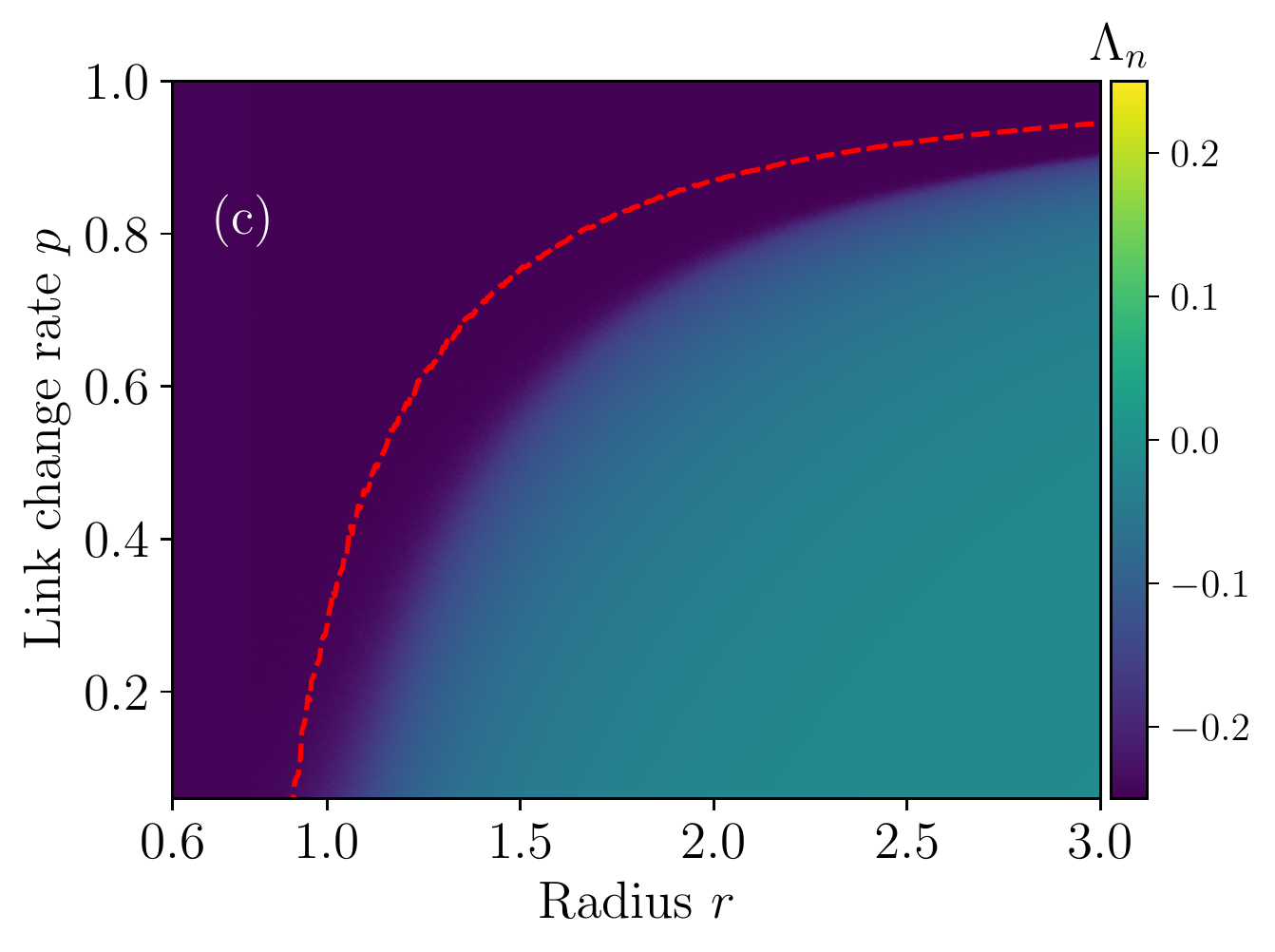}
\includegraphics[width=2.2 in]{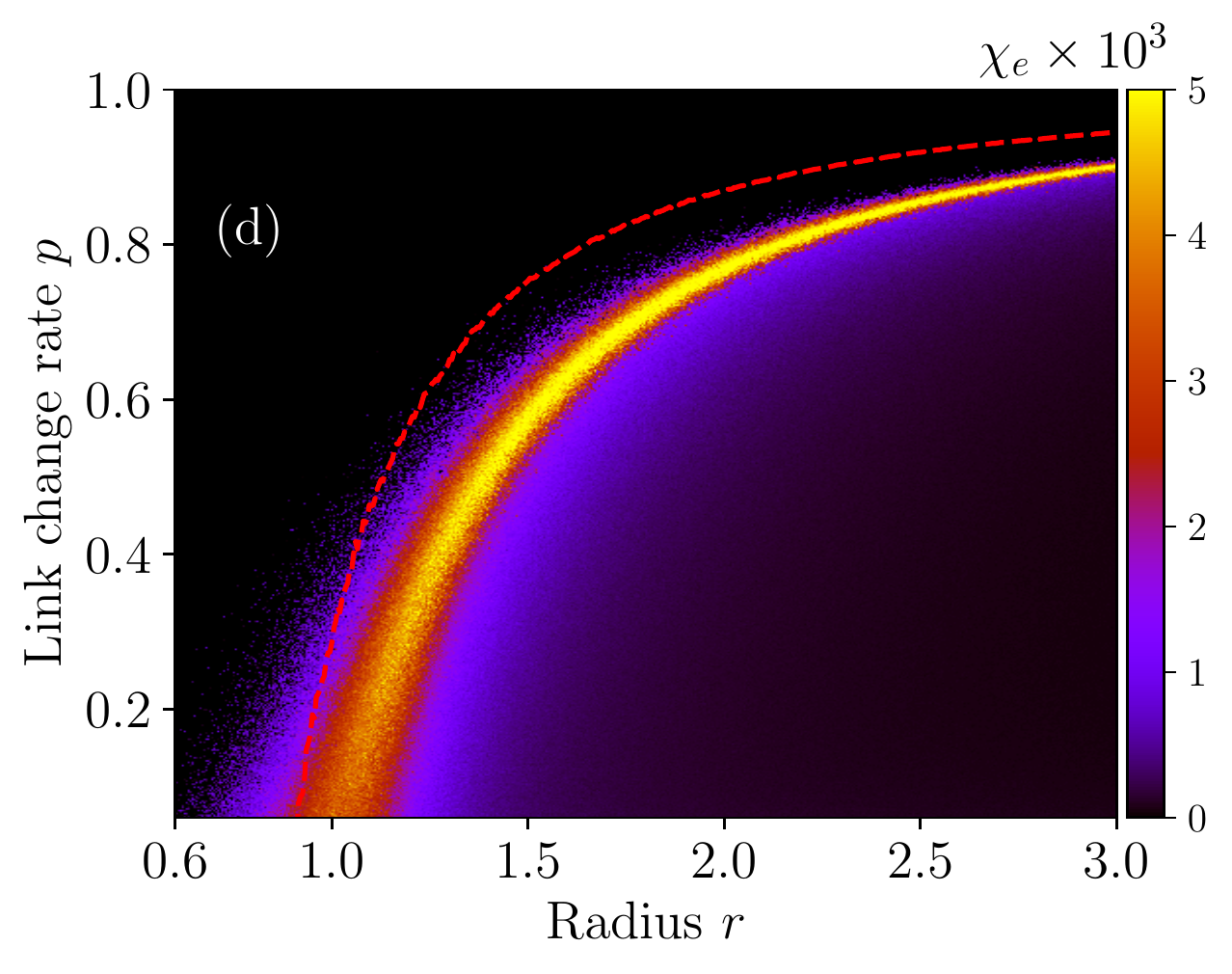}
\includegraphics[width=2.2 in]{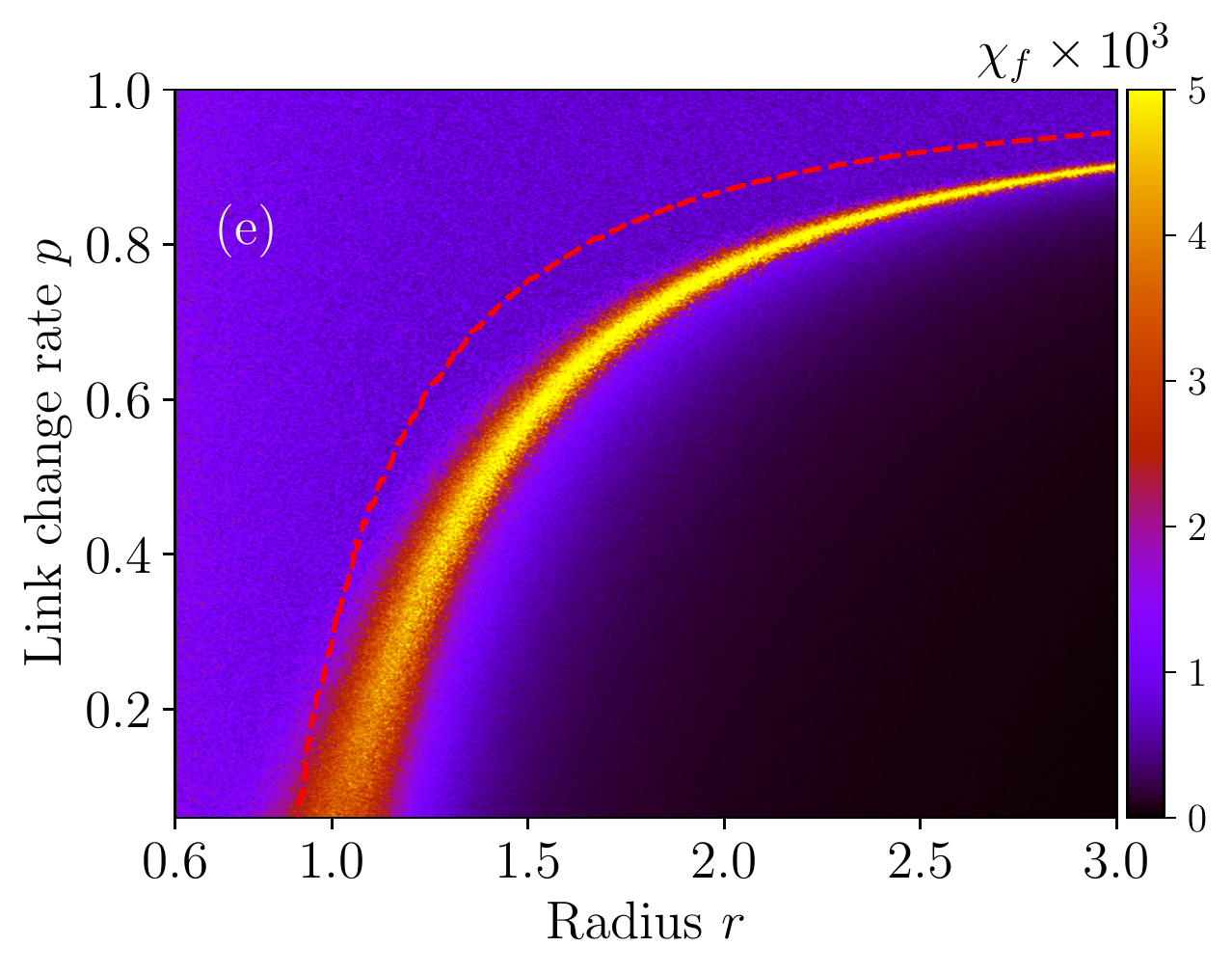}
\includegraphics[width=2.2 in]{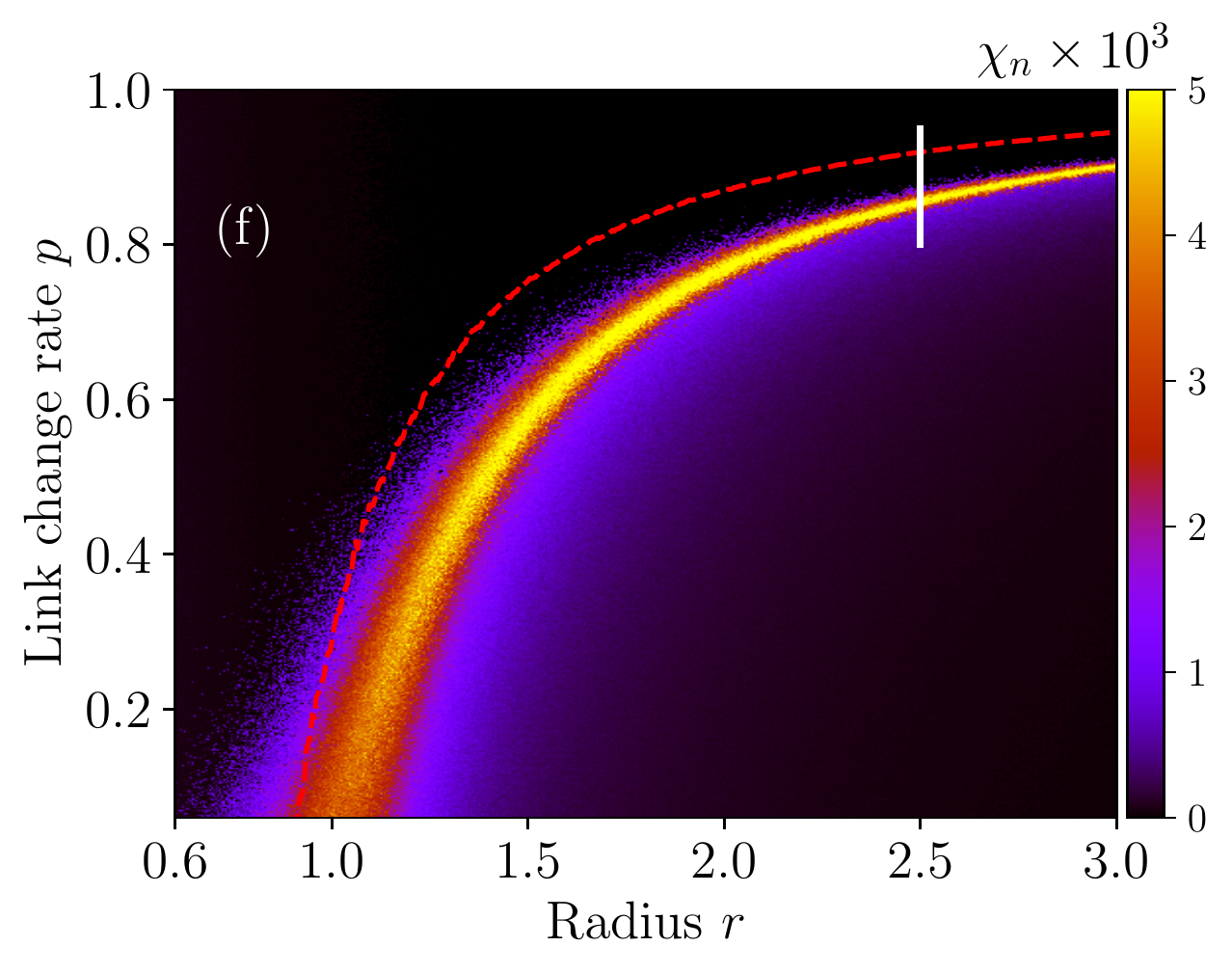}
\caption{Phase diagram of the model on the space $(r,p)$ for three different order parameters. The absorbing phase is on the top left of the region with maximum error and the active one is on the bottom right. Dashed red line is the curve from Eq. (\ref{criticallinered}). Parameters: $M=16^2$, $\phi_0=\ell_0=0.5$, $B=2000$ and $D=100$. There are $501 \times 471 = 235971$ points on each color plot and the respective steps are $\Delta r= 0.005$ and $\Delta p=0.002$.  (a) Density $\rho_e$. The phases are indicated. (b) $\Lambda_f$. (c) $\Lambda_n$. (d) $\chi_e$: error of $\rho_e$. (e) $\chi_f$: error of $\Lambda_f$. (f) $\chi_n$: error of $\Lambda_n$. The white line is the selected region analyzed on Figs. ~\ref{components_domains}.}
\label{phase_diagram}
\end{figure}

The phase diagram on the space $(r,p)$ is shown in Figs. \ref{phase_diagram}(a), \ref{phase_diagram}(b) and \ref{phase_diagram}(c). A large $p$ value (many updates on the link states) combined with low connectivity (low $r$, less interacting neighbors) enable the system to easily get in the absorbing state. The red dashed line indicates the function $p_c$ from Eq. (\ref{criticallinered}). All figures show that the three order parameters have visually similar phase transition curves, different from the red line $p_c(K)$. This happens because Eq. (\ref{criticallinered}) shows the region where $\rho_e$ becomes different from zero on the mean-field approximation while the border between the two regions on the phase diagram indicates where the order parameter becomes non-negligible.

Figures \ref{phase_diagram}(d), \ref{phase_diagram}(e), and \ref{phase_diagram}(f) show the phase diagrams of the respective standard errors of the panels above. The three standard errors have the same behavior except $\chi_f$ in the absorbing phase, showing larger fluctuations. As hinted by the top panels, all three errors are maximum on the same region: the borders on the top figures. These maxima define the location of the phase transition on the space $(r,p)$. 

\subsection{Topological transition: domains and components}

The domain and component-related quantities ($n_d$, $n_c$, $s_d$, and $s_c$) are calculated for each associated network as function of the control parameter $p$ for different network sizes $M$. The idea is to eliminate the effect of the RGG network itself on these quantities, by keeping the radius constant.

\begin{figure} 
\centering
\includegraphics[width=2.3 in]{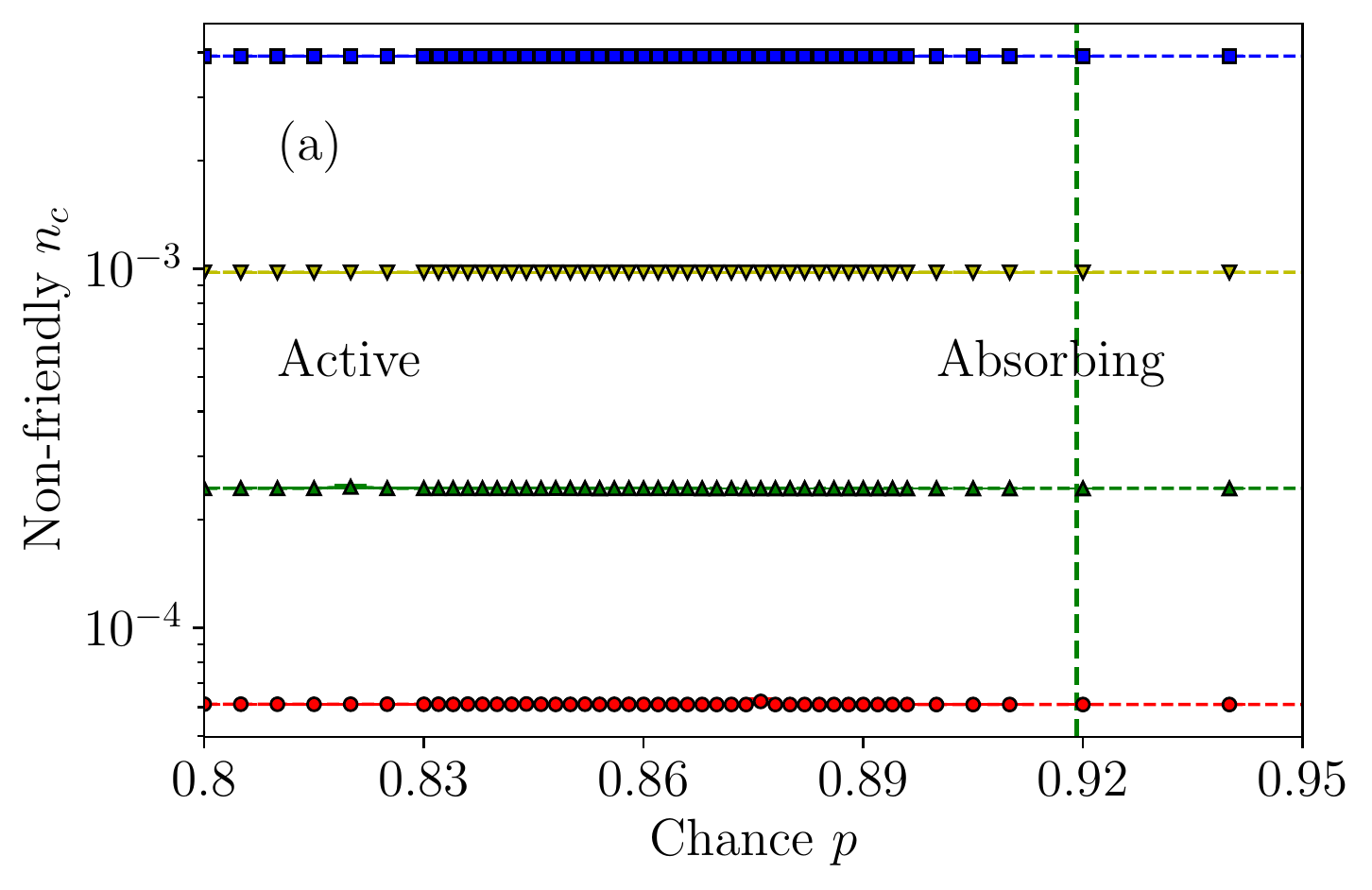}
\includegraphics[width=2.3 in]{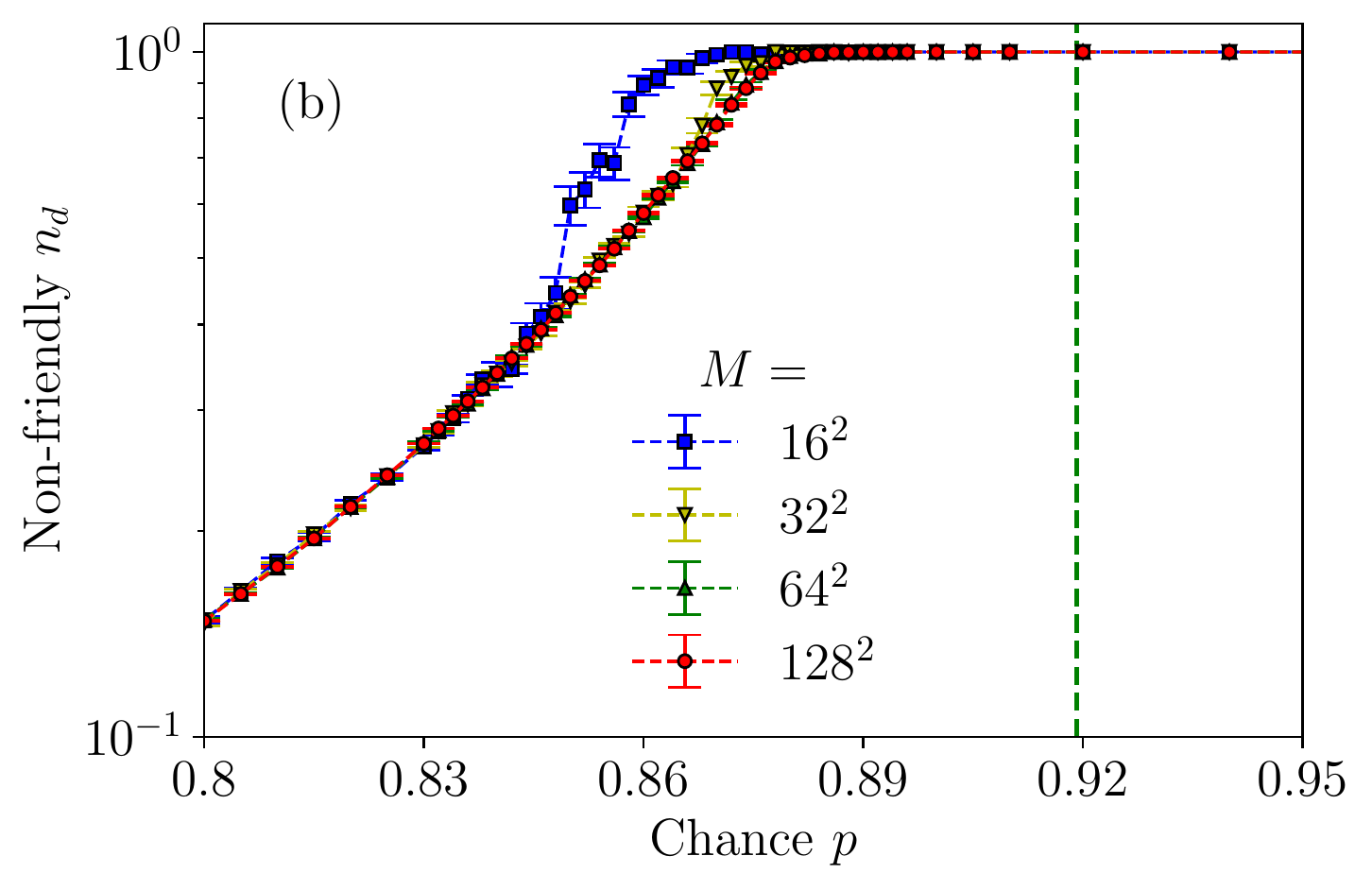}
\includegraphics[width=2.3 in]{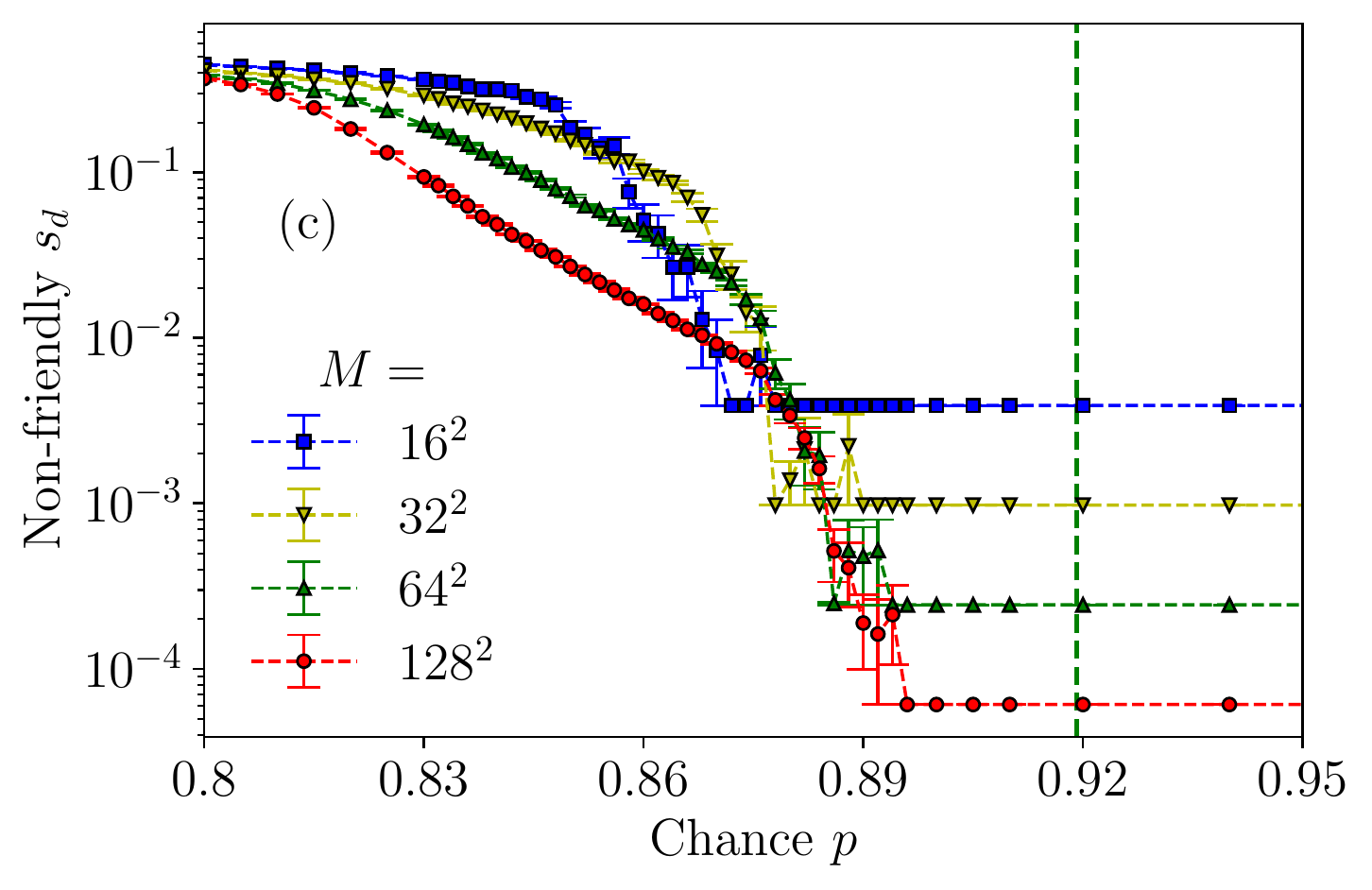}
\includegraphics[width=2.3 in]{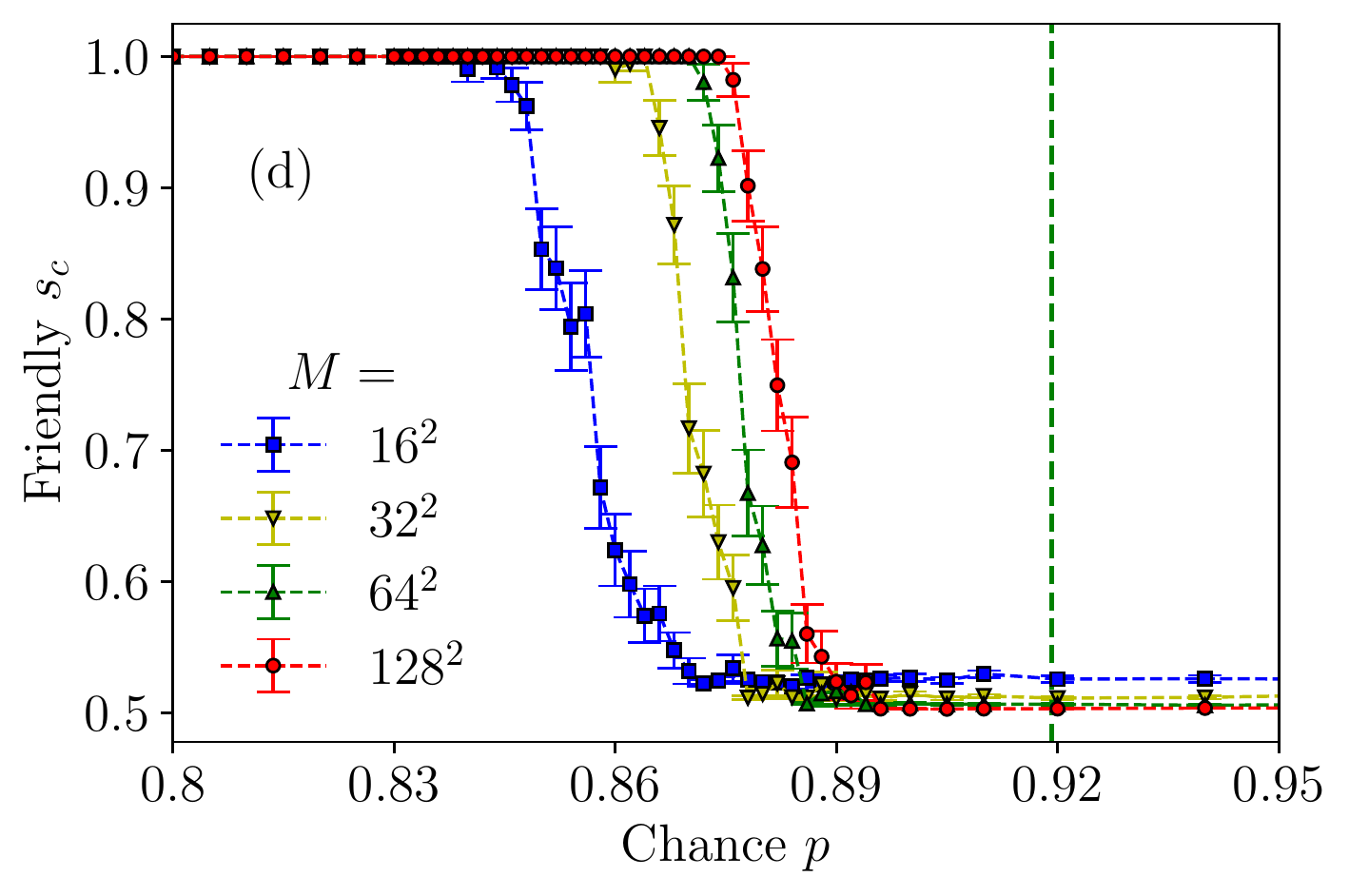}
\includegraphics[width=2.3 in]{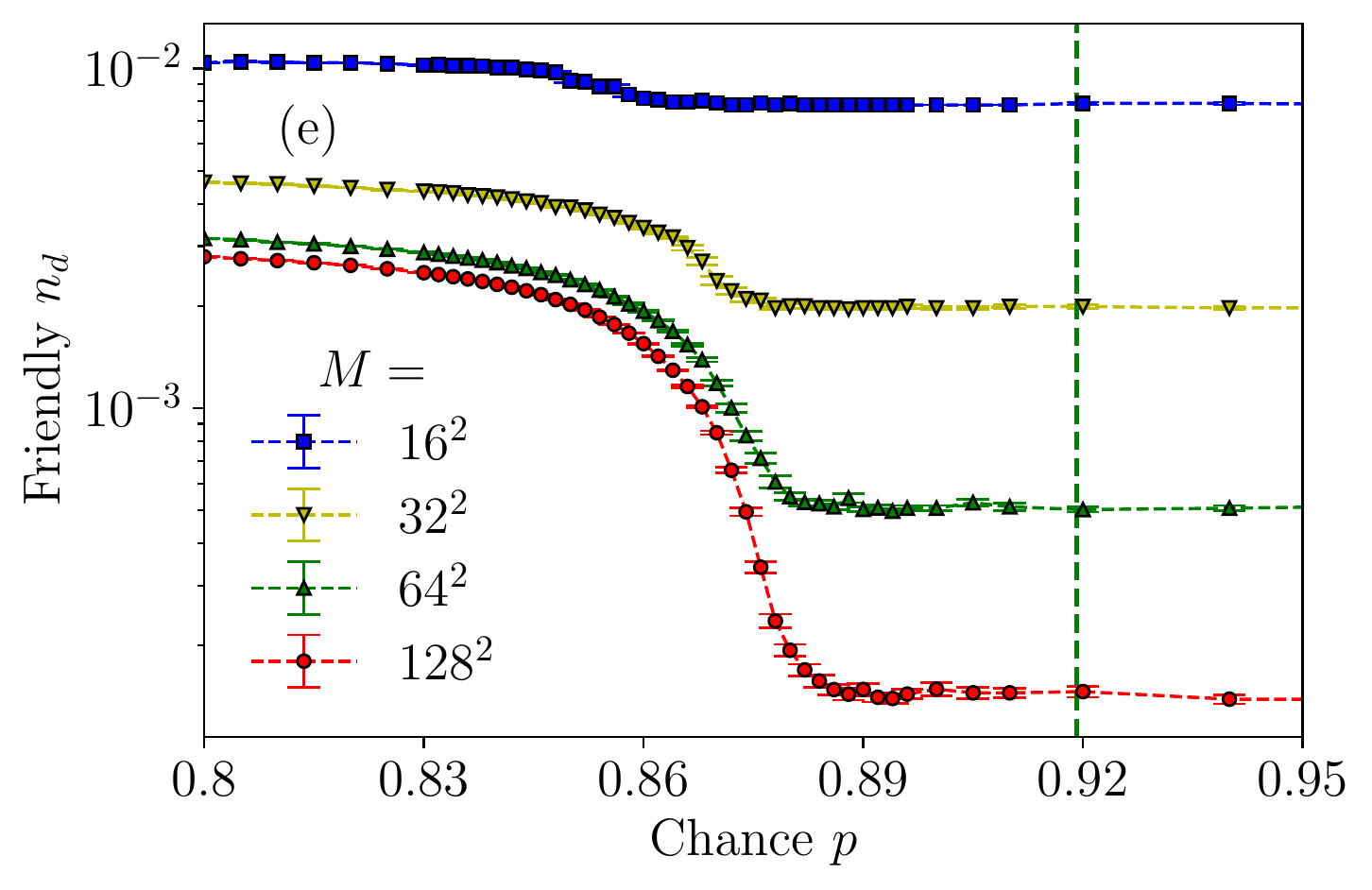}
\includegraphics[width=2.3 in]{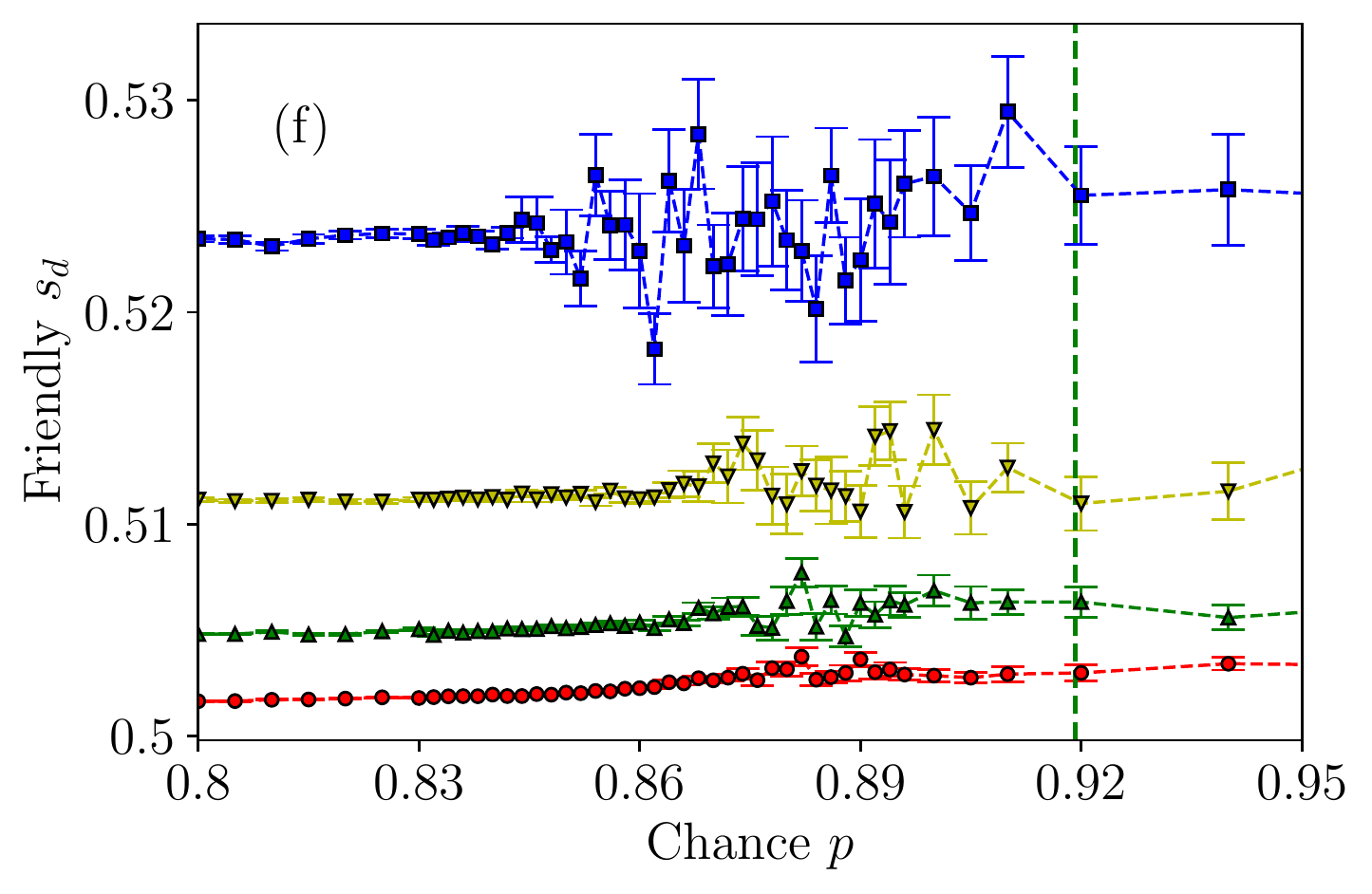}
\caption{Domains and components as function of $p$ for different sizes $M$. Top row: non-friendly network: (a) number of components $n_c$, (b) number of domains $n_d$, (c) size of largest domain $s_d$. Bottom row: friendly network: (d) size of largest component $s_c$, (e) number of domains $n_d$, (f) size of largest domain $s_d$. Parameters: $Q=2000$, $\phi_0=\ell_0=0.5$, $B=500$, $D=10^2$, and $r=2.5$ (which implies an average degree $K \approx 19.6$). The analyzed region is the vertical while line indicated at Fig.~~\ref{phase_diagram}(f). The color scheme is the same for all figures: $M=16^2$, $32^2$, $64^2$, and $128^2$ indicated by blue squares, yellow upside down triangle, green triangle and red circle. The vertical green dashed line indicates $p_c \approx 0.9192$ given by Eq. (\ref{criticallinered}). The active phase occurs on the lower $p$ range while the absorbing one happens on the large $p$ range. All data are approximately constant on the intervals $p<0.8$ and $p_c <  p<1.0$.} 
\label{components_domains}
\end{figure}

Let's begin with the non-friendly network. Figure ~\ref{components_domains}(a) shows that the (normalized) number of components $n_c$ of the non-friendly network does not depend on $p$. Indeed the number of components $n_c$ does not depend on the dynamics, only on the RGG network radius, which does not change with time for each simulation. The size of the largest component for this non-friendly network (not shown) does not have significant change either. On the other hand the (normalized) number of domains $n_d$ depends on the phase, as shown in Fig. \ref{components_domains}(b): in the active phase, $n_d$ increases with $p$ up to 1 for all network sizes and it remains equal to one on the absorbing phase. $n_d =1.0$ means that all (non-friendly) links connect different type of nodes, such that each node is a single domain of size one, as can be seen in Fig. \ref{redes_inativas_ativas}(c). However, on the active phase, the non-friendly links also connect nodes of same state, so when compared to the absorbing phase the domains are larger (Fig. \ref{components_domains}(c)) resulting in a decrease of the number of domains $n_d$. 

In contrast, we can see an effect of the control parameter $p$ on the components of the friendly network. For low $p$ values, on the active phase, the largest component has approximately same size as the associated network: $s_c \approx 1.0$ (Fig.~\ref{components_domains}(d)). This means that the (associated) network has only one component (as shown in Fig.~\ref{redes_inativas_ativas}(e)). As $p$ increases toward the absorbing phase, the unsatisfying pair density decreases creating isolated satisfying pairs (as it is possible to identify in Fig. \ref{components_domains}(d)), which breaks the network in more than one component. This process makes $s_c$ fall to around 0.5, keeping this value throughout the absorbing phase. This value, 0.5, is possibly related to the initial density of friendly links $\ell_0=0.5$, even though more tests are needed to confirm this assertion. The fall of each curve marks the phase transition, and it moves towards the mean field result in the thermodynamic limit, as expected. 

The effect of the friendly networks on the domains is less dramatic than those on the non-friendly network since, although the components get fragmented, the domains retain their size (Fig. \ref{components_domains}(f)). In the absorbing phase the friendly links connect same node states, which limits the size of the domains even if all of them are in the same component. However, on the active phase, only one type of unsatisfying pair appears: type $e$, which is not enough to significantly change this scenario. The average degree $K \approx 19.6$ can contribute to this low impact of pair of type $e$. Probably, in a network with lower average degree, there would be a more significant difference on $s_d$ for the two phases. The motivation for the chosen value $r=2.5$ for these figures is because the phase transition border is almost perpendicular to the $r$ constant line at this value (see white line on Fig.~\ref{phase_diagram}(f)). For a smaller radius the phase transition border is oblique to the $r$ constant line, so the effect of changing $p$ would not be significant.

The number of domains $n_d$ shows a drop when going from the active phase to the absorbing one, which increases with the system size $M$, as shown in Fig. \ref{components_domains}(e). This drop is a direct consequence of the removal of type $e$ pairs. This can be a hint that at the absorbing phase the domains become microscopic in the thermodynamic limit. Even though the largest domain size remains constant through the phase transition. The noticeable features are the large fluctuation on the phase transition. We can understand the same values of $s_d$ on the two phase because the satisfying pairs are both present on these phase. They are the responsible for the size of the largest domain. Once going from active to absorbing, unsatisfying pairs disappear but the satisfying do not, keep the largest domain almost intact.

The topological transition happens on the associated networks because, for example, if a friendly link becomes non-friendly, the friendly network loses one link and the non-friendly one gains one link. There is a real topological transition on the non-friendly network between the two phases of the system, as can be seen on Figs. \ref{components_domains}(b) and (c). The same transition is observed on the friendly network, as can be seen on Figs. \ref{components_domains}(d) and (e). Summarizing, a topological transition does not occur for the complete network but for friendly and non-friendly associated ones.

\subsection{Snapshot of the networks}

\begin{figure}
\centering
\includegraphics[width=6.5 in]{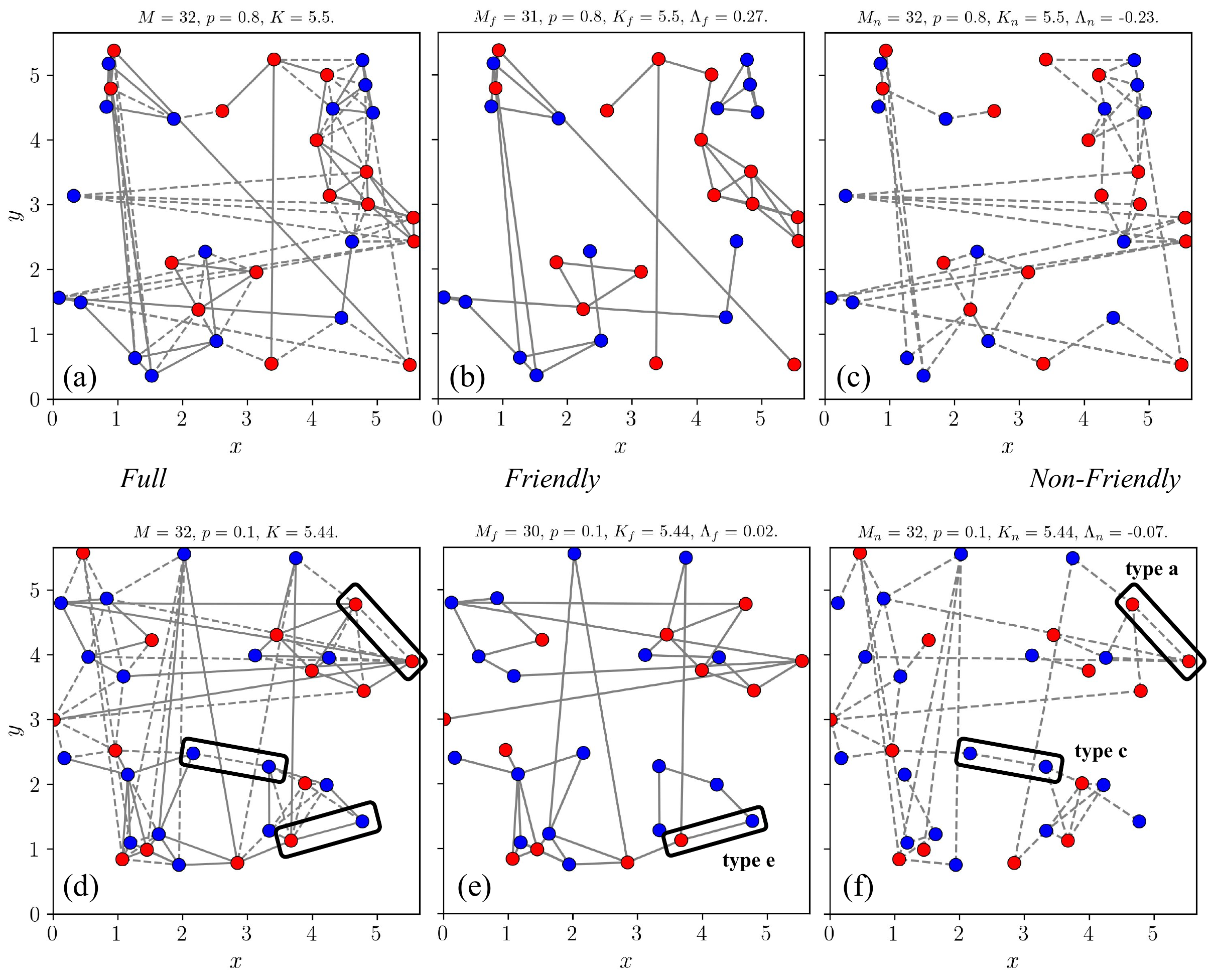}
\caption{Snapshot of the RGG network on the active phase with $M=32$ nodes with linear size of the square box $\sqrt{M} \approx 5.66$ and radius $r=1.4$. The node color (blue or red) indicates its state and solid and dashed gray lines represent friendly and non-friendly links, respectively. For an easy identification, all parameters are indicated on top of each figure. Upper panels (a), (b) and (c) show the absorbing phase with $p = 0.8$ while lower panels (d), (e) and (f) show the active phase with $p = 0.1$. $M$ and $K$ refer to the full network. $\Lambda_f$, $\Lambda_n$, $M_f$, $M_n$, $K_f$, and $K_n$ are the assortativity coefficient, the number of nodes and average degree of the friendly ($f$) and non-friendly ($n$) associated network. Figures (a) and (d) represent full networks while (b) and (e) are related only to friendly associated networks, and (c) and (f) refers to non-friendly associated networks. Isolated nodes do not appear on the associated networks. The dark rectangles highlight unsatisfying pairs at the active phase.} 
\label{redes_inativas_ativas}
\end{figure}

Figure \ref{redes_inativas_ativas} shows multiple snapshots of the full and the associated networks on the two phases of the model. On the absorbing phase, the full network (Fig. \ref{redes_inativas_ativas}(a)) has only satisfying pairs, so that friendly links (solid gray lines) only connect the same type of nodes and non-friendly links (dashed gray lines) only connect different type of nodes. This can be better visualized on the associated networks. Figure \ref{redes_inativas_ativas}(b) shows only the friendly links from Fig. \ref{redes_inativas_ativas}(a). It is easy to see that all pairs have the same type of nodes. On the other hand, Fig. \ref{redes_inativas_ativas}(c) shows the non-friendly associated network containing only the non-friendly links, where one can clearly see that all pairs have different type of nodes.

On the other hand, on the active phase all types of pairs exist. Figure \ref{redes_inativas_ativas}(d) shows the full network in this phase and one can see that all kind of pairs are present. The round rectangles indicate one occurrence of each unsatisfying pair type, certifying that the phase is indeed the active one. In Fig. \ref{redes_inativas_ativas}(e), we present the friendly associated network where the unsatisfying highlighted pair is of type $e$  (see Fig. \ref{Dinamica}). It is easy to see at least six other $e$-type pairs and this is the only unsatisfying type in this friendly network. Figure \ref{redes_inativas_ativas}(f) shows the non-friendly network, where one example of the other two unsatisfying types ($a$ and $c$) are highlighted. 

\section{Conclusions} \label{sec:conclusions}

In summary, we have performed steady-state Monte Carlo simulations in order to investigate the topological transition of the associated networks in a coupled dynamics of node and link states implemented in random geometric graphs. We borrowed the common metrics components and domains used on cultural dissemination to show this transition on each associated network from one phase to the other of the system. We also showed that the modularity (or assortativity coefficient) for the associated networks (friendly and non-friendly) can be used as an order parameter, resulting in the same phase diagrams to that obtained when considering the unsatisfying pair density $\rho_e$. We identified the phase transition as the region with maximum fluctuation of the order parameters. Additionally, the modularity, components and domains show an intuitive picture of the of the associated networks features between the two phases. 

Perspectives of future works include a more detail study of the phase transition itself, determine critical exponents of the model and possibly its universality class. In this sense, we also intend to understand whether the transition causes a space fragmentation, and how to characterize it.

\section*{CRediT authorship contribution statement}

P. F. Gomes: Conceptualization, Methodology, Software, Formal analysis, Investigation, Resources, Writing - Original Draft, Writing - Review Editing. H. A. Fernandes: Methodology, Writing - Review Editing, Formal analysis, Investigation, Resources, Visualization. A. A. Costa: Methodology, Writing - Review Editing, Formal analysis, Investigation, Resources, Visualization, Supervision.

\section*{Declaration of competing interest}
The authors declare that they have no competing financial interests or personal relationships that could have appeared to influence the work reported in this paper.

\section*{Acknowledgments}

This work was supported by FAPEG (P.F. Gomes under the grant 2019 / 102 670 00139). Research was carried out using the computational resources of LaMCAD/UFG and the Center for Mathematical Sciences Applied to Industry (CeMEAI) funded by FAPESP (grant 2013/07375-0).

\appendix
\section{Appendix}
\label{appendixA}
Table \ref{listofsymbols} lists the main parameters used in this work for clarity purposes.

\begin{table}[ht!]
\centering
\begin{tabular}{c|c|c}
\hline \hline
Parameter &  Description & Range    \\
\hline
\hline $B$ & number of MCS for the average calculation, I. & $[1,\infty)$ \\
\hline $D$ & number of network samples for the error calculation, I. & $[1,\infty)$ \\
\hline $\ell(t)$ & density of friendly links at time $t$ $(\ell_0)$ & $[0,1]$ \\
\hline $\ell_0$ & chance of a link be friendly at $t=0$, I. & $[0,1]$ \\
\hline $L$ & number of links & $[1,\infty)$ \\
\hline $\Lambda$ & modularity, Eq. (\ref{lkjweriuowieur}) & $[-1,1]$ \\
\hline $K$ & network average number of neighbors & $[0,M-1]$ \\
\hline $M$ & number of sites of the network, I. & $[1,\infty)$   \\
\hline $\mathcal{N}_c$ & number of components & $[1,M]$ \\
\hline $\mathcal{N}_c$ & number of domains & $[1,M]$ \\
\hline $n_c$ & density of components & $(0,1]$ \\
\hline $n_d$ & density of domains & $(0,1]$ \\
\hline $p$ & link update rate, I. & $[0,1]$ \\
\hline $Q$ & number of MCS in the thermalization, I. & $[1,\infty)$ \\
\hline $r$ & RGG radius, control parameter, I. & $[0,\infty)$ \\ 
\hline $\mathcal{S}_c$ & size of the largest component & $[1,M]$ \\
\hline $\mathcal{S}_d$ & size of the largest domain & $[1,\mathcal{S}_c]$ \\
\hline $s_c$ & density of the size of the largest component & $(0,1]$ \\
\hline $s_d$ & density of the size of the largest domain & $(0,1]$ \\
\hline $\phi (t)$ & density of red nodes at time $t$. & $[0,1]$ \\
\hline $\phi_0$ & chance of an agent be red at $t=0$, I. & $[0,1]$ \\
\hline $t$ & time measured in MCS & $[1,Q+B]$ \\
\hline $\chi$ & standard error  & $(0,\infty)$ \\
\hline
\hline
\end{tabular}
\caption{List of the most used  symbols on this work. The letter I indicates that the quantity is an input parameter on the Fortran code.}
\label{listofsymbols}
\end{table}

\end{document}